\documentclass[aps,preprint,a4paper,showabstract,superscriptaddress]{revtex4-2}

\usepackage{multirow}
\usepackage{graphicx}
\usepackage{epstopdf}
\usepackage{dcolumn}
\usepackage{bm}
\usepackage{hyperref}
\usepackage[utf8]{inputenc} 
\usepackage[T1]{fontenc}    
\usepackage{lmodern}
\usepackage{url}            
\usepackage{booktabs}       
\usepackage{amsfonts}       
\usepackage{nicefrac}       
\usepackage{microtype}      
\usepackage[version=3]{mhchem}
\usepackage{natbib}
\usepackage{amsmath}
\usepackage{amssymb}
\usepackage{xcolor}
\usepackage[range-units = single,range-phrase = \text{--},separate-uncertainty = true,detect-all]{siunitx}
	\DeclareSIUnit\linepair{lp}
	\DeclareSIUnit\pixels{px}
\usepackage{soul}
\setstcolor{red}
\usepackage{cancel}
\def\bibsection{\noindent \textbf{ References.}}

\renewcommand\labelenumi{(\roman{enumi})}
\renewcommand\theenumi\labelenumi

\begin{document}

\title{Physics-Based Learning of the Wave Speed Landscape in Complex Media}

\author{Baptiste~H\'eriard-Dubreuil}
\thanks{Corresponding authors. Email: baptiste.heriard-dubreuil@espci.fr, alexandre.aubry@espci.fr}
\affiliation{Institut Langevin, ESPCI, PSL University, CNRS, Paris, France.}
\author{Emma~Brenner}
\affiliation{Institut Langevin, ESPCI, PSL University, CNRS, Paris, France.}
\affiliation{SuperSonic Imagine Company, Aix-en-Provence, France.}
\author{Benjamin~Rio}
\affiliation{Institut Langevin, ESPCI, PSL University, CNRS, Paris, France.}
\author{William~Lambert}
\affiliation{SuperSonic Imagine Company, Aix-en-Provence, France.}
\author{Foucauld~Chamming's}
\affiliation{Institut Bergoni\'e, Bordeaux, France.}
\author{Mathias~Fink}
\affiliation{Institut Langevin, ESPCI, PSL University, CNRS, Paris, France.}
\author{Alexandre Aubry}
\thanks{Corresponding authors. Email: baptiste.heriard-dubreuil@espci.fr, alexandre.aubry@espci.fr}
\affiliation{Institut Langevin, ESPCI, PSL University, CNRS, Paris, France.}

\date{\today}
\begin{abstract} \bfseries \boldmath
   The propagation of waves through complex, inhomogeneous media is a fundamental challenge across scales, from optics to geophysics. In such environments, the waves reflected by the medium are typically dominated by a random speckle pattern, a complex interference resulting from multiple scattering by fine-scale heterogeneities. While this speckle is traditionally viewed as a source of noise that blurs the medium's structure, it contains rich information about the underlying wave speed landscape. Here, we introduce a physics-based learning framework that reconstructs the velocity distribution from the reflection matrix of the medium. We show how the speckle wave-field can be exploited to probe the focusing quality, serving as an intrinsic guide star to evaluate the local consistency of the velocity model. By representing wave propagation as a differentiable multi-layer network, we leverage optimization tools to solve the inverse problem and infer the wave speed map. We validate this approach through ultrasound experiments in highly scattering tissue-mimicking phantoms and human breast tissues, demonstrating its potential for tumor detection and characterization where standard methods fail. This synergy between wave physics and gradient-based optimization provides a general paradigm for non-invasive tomography in any complex environment where reflection data is available.
\end{abstract}
\maketitle
\newpage

\noindent {\large \textbf{Introduction}}

As waves propagate through heterogeneous media, they accumulate complex phase and amplitude perturbations~\cite{Bertolotti2022}. Large-scale variations in the phase velocity induce wavefront distortions~\cite{Judkewitz2015}. This phenomenon is exploited in transmission-based diffraction tomography across optics~\cite{Wolf1969}, ultrasonics~\cite{Devaney1982}, and seismology~\cite{Thurber2015} to map the underlying velocity landscape. However, the vast majority of in vivo or in situ applications are restricted to epi-detection configurations, where one must rely solely on backscattered echoes generated by fine-scale heterogeneities. In these reflection modes, the recorded waves primarily encode the medium's reflectivity rather than its constitutive wave speed, typically yielding structural maps that are qualitative in nature. Recovering the quantitative velocity distribution from such reflection data remains a fundamental challenge. Solving it would transform wave-echo imaging from a qualitative structural tool into a quantitative probe of a medium's physical properties.

Recovering velocity variations in media characterized by both large-scale fluctuations and fine-scale heterogeneities is a non-trivial inverse problem. This challenge has encouraged the development of computational frameworks that move beyond the single-scattering (Born) approximation to exploit multiple scattering, both in transmission~\cite{kamilov2015learning, kamilov2016optical, chen2020multi} and reflection~\cite{virieux2009overview, Li2025, wasik2026}. However, these methods typically require the simultaneous reconstruction of all scales of heterogeneities, a task that is computationally intensive, sensitive to model mismatch, and prone to convergence issues. Yet, many real-world environments contain sub-wavelength features that give rise to complex, random interference patterns known as speckle. In fields such as medical ultrasound, reflection seismology and optical coherence tomography, where sub-wavelength heterogeneities dominate the back-scattered wave-field, the deterministic inversion of these fine-scale features becomes fundamentally ill-conditioned. In this speckle regime, standard inversion frameworks fail, effectively preventing the estimation of the wave velocity landscape.

To recover large-scale velocity variations in such complex media, we adopt an alternative strategy that decouples wavefront distortions induced by the background velocity landscape from the stochastic scattering generated by fine-scale heterogeneities (Figs.~\ref{fig:split_step}a-d). While preliminary versions of this decoupling have been explored in ultrasound via straight-ray approximations~\cite{stahli2020improved, Simson2024, simson2025ultrasound} or multi-layered geometries~\cite{ali2021local, heriard2023refraction}, we propose a more rigorous physical treatment rooted in matrix imaging~\cite{Yoon2020, badon_distortion_2020, bureau2023three}. This generic framework operates on the reflection matrix of the medium, manipulating the synthetic focusing of both emitted and backscattered waves in post-processing. By leveraging speckle statistics~\cite{lambert2020reflection} and the associated memory effect~\cite{Osnabrugge2016}, the matrix approach provides a direct measure of the local focusing quality~\cite{lambert2020reflection}. It thus reveals the wavefront distortions accumulated during propagation, essentially using the random scattering as an intrinsic distribution of guide stars to probe the medium's hidden aberrations~\cite{lambert_distortion_2020, najar_non-invasive_2023, bureau2023three}.

\begin{figure*}[htp]
	\centering
	\includegraphics[width=\textwidth]{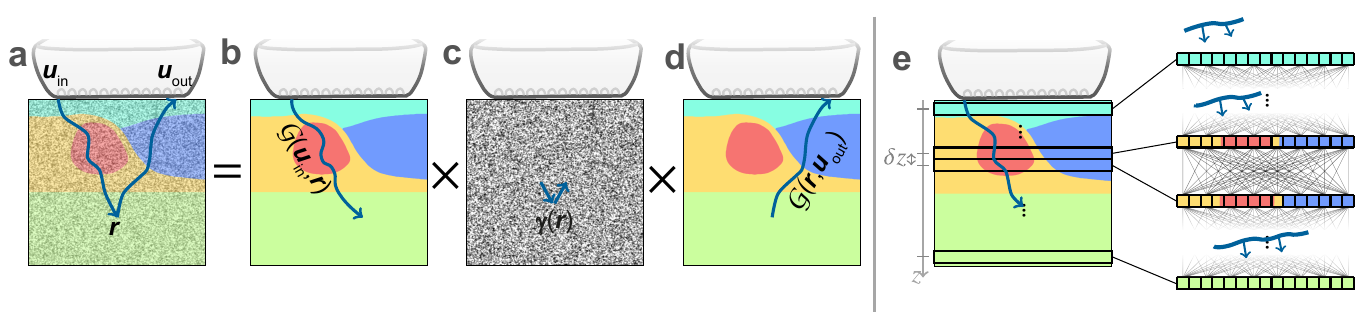}
	\caption{\label{fig:split_step} \textbf{Model of wave propagation in a multi-scale heterogeneous medium.} \textbf{a}-\textbf{d}, Wave propagation in a reflection configuration (\textbf{a}) can be modelled as the succession of a forward propagation (\textbf{b}), a single back-scattering event caused by short-scale heterogeneities (B\&W map) (\textbf{c}), and a back-propagation (\textbf{d}). The forward (\textbf{b}) and backward (\textbf{d}) propagation are driven by the large-scale variations of wave velocity (colored map). \textbf{e}, These unidirectional paths include forward multiple scattering events that can be modelled as a succession of diffraction phase screens (colored arrays) and free-space operators (connections) in homogeneous layers of thickness $\delta z$. This forward model takes the form of a multi-layer network.}
\end{figure*}

In this work, we introduce a learning-based formulation of matrix imaging that exploits the quality of wave focusing as an intrinsic guide star. We represent wave propagation as a differentiable multi-layer model, utilizing cascaded diffraction phase screens~\cite{kang_tracing_2023, Haim2024} interleaved with split-step free-space propagation~\cite{hardin1973applications, stoffa_splitstep_1990} (Fig.~\ref{fig:split_step}e). This architecture enables the use of gradient-based optimization to iteratively recover the wave velocity landscape from reflection data alone, thereby breaking the usual limits of inversion approaches in the speckle regime. We validate this approach  through ultrasound experiments on tissue-mimicking phantoms, ex vivo tissues, and in vivo human breast data, achieving quantitative speed-of-sound imaging in a reflection geometry, a configuration where conventional tomography typically fails. Given its reliance on the general reflection matrix framework, this paradigm is inherently universal and can be extended to any wave-based imaging system, from mesoscopic optics to planetary-scale seismology.


\vspace{5mm} 
\noindent {\large \textbf{Probing Focusing Quality in Speckle}}

We consider an imaging configuration in which the reflection matrix is acquired using a transducer array in direct contact with the sample (Fig.~\ref{fig:method}a). The simplest acquisition sequence consists in insonifying the medium with one element at a time and recording with all elements the time-dependent field reflected back from the medium (see Methods for alternative sequence). The measured data (Fig.~\ref{fig:method}b) form a time-dependent reflection matrix, $\mathbf{R}_{\mathbf{uu}}(t)=[\mathcal{R}(\mathbf{u}_{\textrm{in}},\mathbf{u}_{\textrm{out}},t)]$, whose elements encode the impulse responses between each source/receiver pair identified by their position $\mathbf{u}_{\textrm{in/out}}$ (Fig.~\ref{fig:split_step}a). As we make the single backscattering approximation, this matrix can be expressed in the frequency domain as an integral over the medium reflectivity $\gamma(\mathbf{r})$, corresponding to the short scale heterogeneities (Fig.~\ref{fig:split_step}c), convolved by Green’s functions $ \mathcal{G}(\mathbf{u},\mathbf{r},f)$ that account for wave propagation through the large-scale velocity distribution (Figs.~\ref{fig:split_step}b,d):
\begin{equation}
    \mathcal{R}(\mathbf{u}_{\textrm{in}},\mathbf{u}_{\textrm{out}},f)=\int d\mathbf{r} \,  \mathcal{G}(\mathbf{u}_{\textrm{in}},\mathbf{r},f) \gamma(\mathbf{r})  \mathcal{G}(\mathbf{u}_{\textrm{out}},\mathbf{r},f).
\end{equation}
To form an image, the recorded signals are numerically focused at depth using an estimate of the transmission matrix, $\mathbf{T}_{\mathbf{x u}}(z,f;c)=[ \mathcal{T}(\mathbf{x}, \mathbf{u}, z,f;c)]$. This matrix describes wave propagation between the transducer array and each focal plane at depth $z$ under a trial velocity distribution $c(\mathbf{r})$. Its phase conjugate can be used to project the reflection matrix into the focused basis as follows~\cite{lambert2020reflection}: 
\begin{equation}
\label{foc}
    \overline{\mathbf{R}}_{\mathbf{xx}}(z;c)=\sum_f \mathbf{T}_{\mathbf{xu}}^*(z,f;c) \times \mathbf{R}_{\mathbf{uu}}(f)  \times  \mathbf{T}_{\mathbf{xu}}^{\dag}(z,f;c),
\end{equation}
where the symbols $\times $, $*$ and $\dag$ stand for matrix product, conjugate, conjugate transpose, respectively. In this work, the matrix $\mathbf{T}_{\mathbf{xu}}(z,f;c)$ will be obtained from a velocity hypothesis $c(\mathbf{r})$ using a forward model referred to as the split-step Fourier method in geophysics \cite{hardin1973applications,stoffa_splitstep_1990} or the beam propagation method in optics \cite{feit1978light}.  Under this approach, wave propagation is described as a sequence of diffraction phase screens capturing lateral refractive-index fluctuations, $\delta n(x,z)=n(x,z)-\bar{n}(z)$, interleaved with free-space propagation governed by the laterally-averaged refractive index $\bar{n}(z)=\langle n(x,z)\rangle_x$, with $n(x,z)=c_0/c(x,z)$ (Fig.~\ref{fig:split_step}e, Methods).
Equation~\eqref{foc} yields focused reflection matrices $  \overline{\mathbf{R}}_{\mathbf{xx}}(z;c)$ (Fig.~\ref{fig:method}d), whose diagonal elements correspond to confocal signals conventionally used to estimate the medium reflectivity (Fig.~\ref{fig:method}e).
Importantly, the off-diagonal structure of these matrices provides additional information: the spreading of backscattered energy away from the diagonal directly reflects wavefront distortions induced by mismatches between our wave speed hypothesis $c(\mathbf{r})$ and the actual velocity distribution within the numerical aperture of our imaging system~\cite{lambert2020reflection}.

This property allows us to define a focusing quality metric that quantifies the efficiency of numerical focusing at each depth. Interestingly, we can define such a metric without knowing the actual reflectivity $\gamma(\mathbf{r})$, simply by measuring the average energy spreading outside of the main diagonal. Specifically, we compute the averaged ratio $\mathcal{F}$ between the confocal energy and the total backscattered energy in $\overline{\mathbf{R}}_{\mathbf{xx}}(z;c)$ (Fig.~\ref{fig:method}f):
\begin{equation}
    \mathcal{F}(c) = \left\langle \frac{|\overline{\mathcal{R}}(\mathbf{x}, \mathbf{x}, z; c)|^2}{\sum\limits_{\mathbf{x'}} |\overline{\mathcal{R}}(\mathbf{x}', \mathbf{x}, z; c)|^2} \right\rangle_{\mathbf{x}, z}.
    \label{eq:focusing_quality2}
\end{equation}
The symbol $\langle \cdots \rangle$ stands for an average over the variables in subscript. This dimensionless quantity serves as a physically meaningful indicator of model adequacy: when the velocity model matches the true medium, focusing is optimal and the metric is maximized.

\begin{figure*}[ht]
	\centering
	\includegraphics[width=\textwidth]{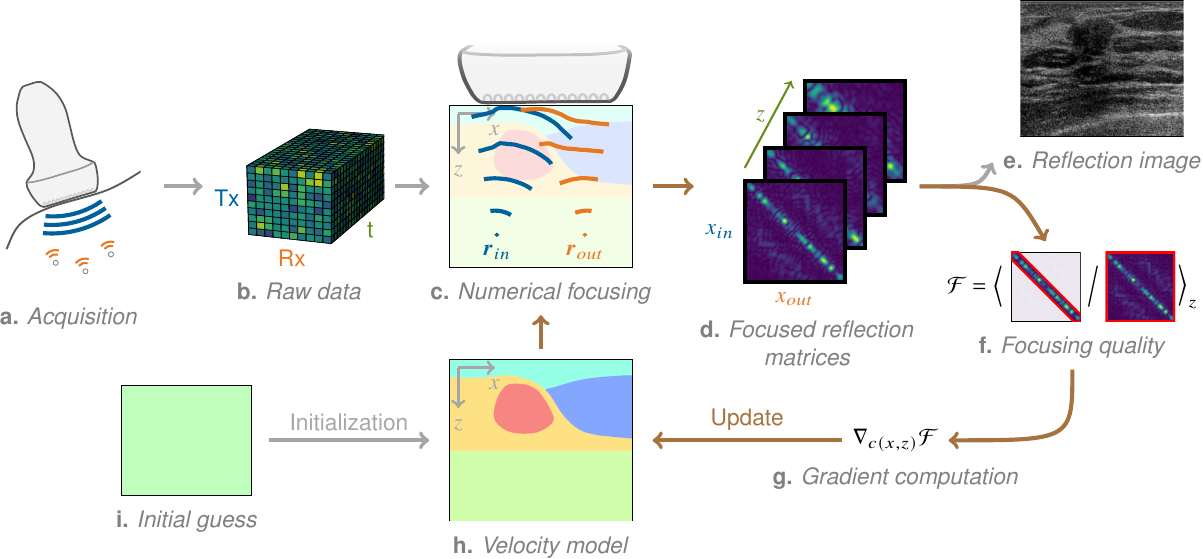}
	
	\definecolor{morange}{rgb}{0.91, 0.47, 0.13}
	\definecolor{mblue}{rgb}{0., 0.38, 0.61}
	\definecolor{mgreen}{rgb}{0.4, 0.55, 0.1}
	\caption{\label{fig:method} \textbf{Differential matrix imaging.} 
		\textbf{a}, an ultrasound acquisition is performed with an ultrasound probe. \textbf{b}, Raw data is gathered in a 3D tensor, indexed by emission ({\color{mblue}Tx}), reception ({\color{morange}Rx}) and time ({\color{mgreen}$t$}). \textbf{c}, A numerical focusing algorithm focuses the signals at point $\color{mblue}\mathbf{r}_{in}$ in emission and $\color{morange}\mathbf{r}_{out}$ in reception according to a velocity model (\textbf{h}). \textbf{d}, The repetition of this operation for different depths and lateral positions yields focused reflection matrices. From these matrices, we obtain either a reflection image (\textbf{e}) or a focusing quality (\textbf{f}) computed by averaging the energy on the diagonal, divided by the total energy of reflection matrices. \textbf{g}, The focusing quality gradient with respect to the velocity model is then computed to update the current velocity model, in a gradient ascent scheme. \textbf{i}, An initial guess of the velocity model (often chosen as uniform) is required to start the optimization process.}
\end{figure*}

\vspace{5mm} 
\noindent {\large \textbf{Focusing Quality as a Guide Star for Optimization}}

Rather than performing an exhaustive search over velocity models, we exploit the differentiability of the focusing process to perform a gradient ascent of the focusing quality with respect to the velocity distribution (brown loop in Fig.~\ref{fig:method}). To do so, we can compute its derivative using the chain rule:
\begin{equation}
	\nabla_c \mathcal{F}(c) = \sum_{z}\frac{\partial \mathcal{F}(c)}{\partial \overline{\mathbf{R}}_{\mathbf{xx}}(z; c)} \cdot  \frac{\partial \overline{\mathbf{R}}_{\mathbf{xx}}(z; c)}{\partial c}.
	\label{eq:gradient}
\end{equation}
This derivative indicates the direction in which updating the velocity map would result in the highest increase of the focusing quality. This formulation transforms wave velocity estimation into an optimization problem driven by an intrinsic, data-derived loss function, without requiring access to transmission measurements or prior knowledge of short-scale disorder.

The gradient-ascent framework is highly flexible, as each block in Fig.~\ref{fig:method} can be implemented by any fully differentiable operator. This allows the incorporation of advanced focusing strategies that account for refraction, diffraction, and forward multiple scattering.

Thanks to this architecture, our approach can actually be reformulated as a self-supervised learning problem. On the one hand, the focusing quality function plays the role of a loss. On the other hand, the numerical focusing process can be expressed as a multi-layer network made of a succession of elementary operations analogous to neural layers. Their weights are dictated by diffraction and the activation functions are parameterized by the local wave velocity. We can therefore benefit from the development of resources dedicated to the implementation and training of deep neural network. These resources include auto-differentiable languages (e.g. PyTorch \cite{paszke2019pytorch}, JAX \cite{schoenholz2020jax}) designed for the gradient-descent-based training of neural networks. Such languages combine Python's efficient prototyping with built-in optimization features, automatic differentiation and GPU-based accelerations. 

The use of such auto-differentiable languages to learn cascading phase screens has recently been experimented in optics in order to see through a complex medium \cite{Haim2024}. Here, we generalize this idea to the learning of the wave velocity field itself within the complex medium.
To encompass a wide range of heterogeneous media, we use a large number of phase screens (typically one par wavelength), and optimize them all simultaneously. The resulting high dimensional problem is ill-posed and strongly non-linear, rising several convergence issues.
To overcome such challenges, we enforce a slowly spatially varying velocity field, as per the large-scale assumption of Fig.~\ref{fig:split_step}, using a total-variation regularization \cite{rudin1992nonlinear} and a spatial filtering step (see Methods). In addition, we perform a Lagrangian relaxation to split the optimization variables, thus increasing the problem dimension and facilitating the convergence.

The whole post-processing pipeline and hyper-parameters have been tuned and validated by numerical simulations described in Supplementary Section~\ref{sec:simulation}. This numerical study also demonstrates the robustness of our approach with respect to the reflection data quality (Supplementary Section~\ref{sec:acquisition_effect}), in particular, the emission basis and its incompleteness (Supplementary Fig.~\ref{fig:sup_tx}) as well as the signal-to-noise ratio (Supplementary Fig.~\ref{fig:sup_snr}) .

\vspace{5mm} 
\noindent {\large \textbf{From 2D to 3D \textit{Ex Vivo} Imaging}} 

Ultrasound imaging offers a convenient and versatile platform to demonstrate our approach. Using transducer arrays, the reflection matrix can be readily measured by emitting broadband pulses and recording the reflected wave fields with wavelength-scale spatial sampling (Fig.~\ref{fig:method}a; Methods). Soft biological tissues are particularly well suited to this framework, as they combine a random distribution of unresolved scatterers and pronounced sound-speed variations across tissue types such as fat, skin, and muscle \cite{Duck1990}.

\begin{figure*}[htp]
		\centering
		\includegraphics[width=\textwidth]{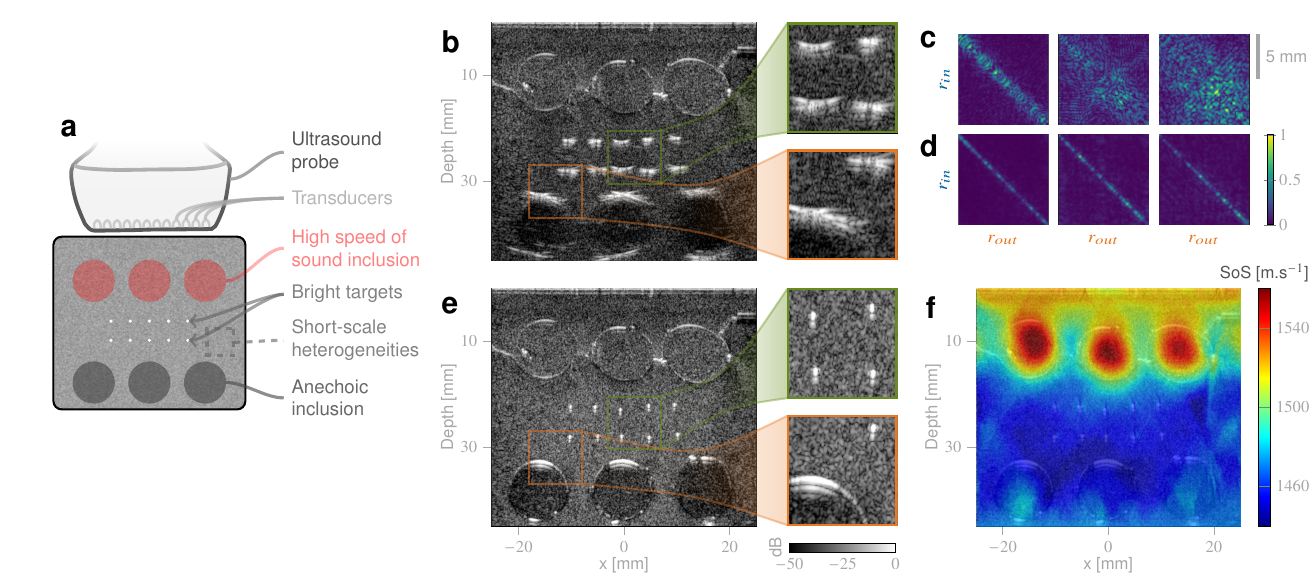}
		
		\caption{\textbf{Ultrasound aberration correction and speed of sound imaging in a 2D \textit{in vitro} experiment.}
			Data obtained from Ref.~\cite{ali2023sound}.
			\textbf{a}, Experiment in which an ultrasound probe is positioned on top of a homemade tissue-mimicking phantom with speed of sound inclusions~\cite{ali2023sound}. \textbf{b},  Aberrated reflection image obtained with a uniform speed of sound hypothesis of $c=1480$~m.s$^{-1}$. Two zoomed sections are displayed on the right. \textbf{c}-\textbf{d},  Examples of focused reflection matrices obtained at initialization (\textbf{c}) and at convergence (\textbf{d}) at depths of 20~mm, 30~mm and 40~mm, normalized by their maximum. \textbf{e}, Corrected reflection image obtained at convergence with the speed of sound map displayed in (\textbf{f}). \textbf{f}, Obtained speed of sound map (overlaid on the corrected reflection image), color coded as a function of the speed of sound expressed in m.s$^{-1}$.}
		\label{fig:2d_invitro}
	\end{figure*}
	
As a proof of concept, we apply our approach to open ultrasound data from a 2D in vitro experiment \cite{ali2023sound} (Fig.~\ref{fig:2d_invitro}a). The tissue-mimicking phantom generates random speckle and contains shallow sound-speed inclusions ($z\sim10$ mm), deep anechoic regions ($z\sim40$ mm), and intermediate-depth nylon rods acting as point scatterers. A conventional image beamformed with a homogeneous sound speed ($c_0=1480$ m·s$^{-1}$) is shown in Fig.~\ref{fig:2d_invitro}b.

Sound-speed heterogeneities induce strong aberrations, visible as distortions of point reflectors and deformations of anechoic interfaces, and as a progressive off-diagonal spreading of the focused reflection matrices with depth (Fig.~\ref{fig:2d_invitro}c). This behavior is quantitatively captured by the focusing-quality metric $\mathcal{F}(c)$ (Eq.~\ref{eq:focusing_quality2}), measured at 11.3\% before correction. Using this metric as a guide, the differential matrix imaging (DMI) optimization (Fig.~\ref{fig:method}) converges towards an optimal velocity field.
The resulting sound-speed map (Fig.~\ref{fig:2d_invitro}f) accurately recovers the expected structure, with well-resolved inclusions exhibiting a $\sim$100 m·s$^{-1}$ contrast, outperforming previous approaches \cite{simson2025ultrasound, schweizer2025pulse}.

After correction, the reflection matrices collapse onto their diagonals (Fig.~\ref{fig:2d_invitro}d), indicating diffraction-limited focusing. We reach a focusing quality of 21.3\%, representing an 88\% increase. The corresponding confocal images (Fig.~\ref{fig:2d_invitro}e) resolve individual point-like scatterers and restore the circular geometry of anechoic regions, demonstrating the benefit of an optimized sound-speed model for reflectivity imaging (Supplementary Video 1).

\begin{figure*}[ht]
	\centering
	\includegraphics[width=\textwidth]{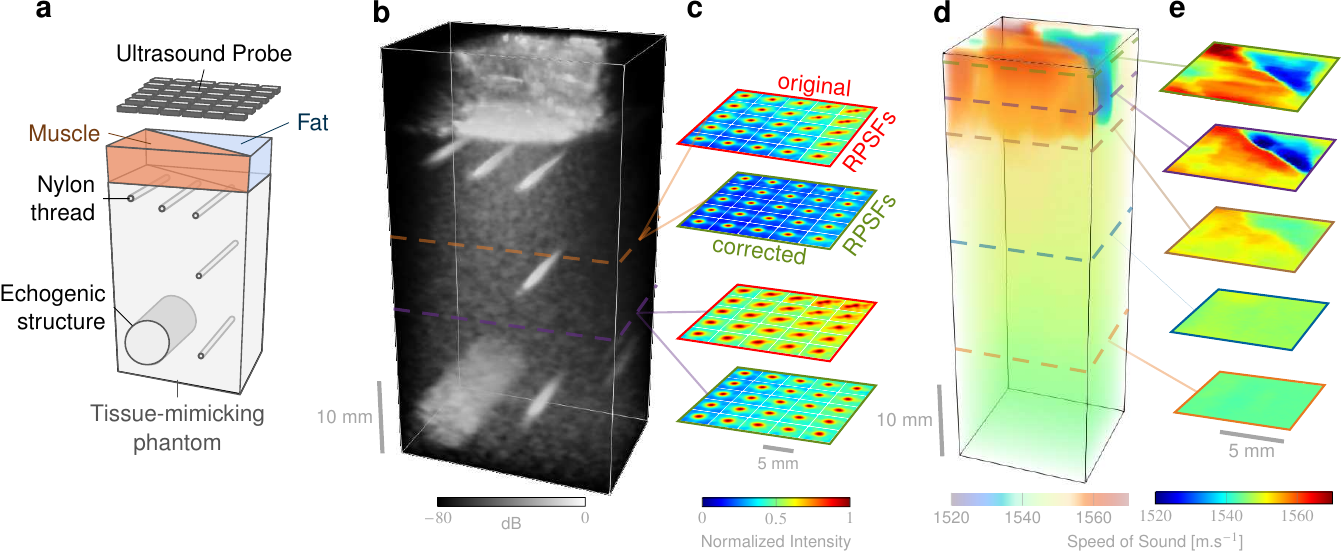}
	
	\caption{\label{fig:3d_invitro} \textbf{Speed of sound imaging of \textit{ex vivo} tissues in a 3D configuration.} 
		Data obtained from~\cite{bureau2023three_dataset}.
		\textbf{a}, Experiment in which a 1024 transducers matrix probe images a tissue-mimicking phantom through a layer of pork tissue of fat and muscle~\cite{bureau2023three}. \textbf{b}, Corrected reflection image obtained with the 3D speed of sound map displayed in (\textbf{d}) (see Supplementary Video 2 for comparison with the original image). \textbf{c}, Examples of focal spots estimated with the focused reflection matrices, at initialization (red) and after correction (green), for depths $z=$30~mm and 40~mm. \textbf{d}, Obtained 3D speed of sound map, color coded as a function of the speed of sound (in~m.s$^{-1}$). \textbf{e}, Horizontal slices of the 3D speed of sound map, at 5~mm, 10~mm, 15~mm, 30~mm and 45~mm. The speed of sound is displayed in m.s$^{-1}$.}
	
\end{figure*}

To demonstrate the extension of our framework to three-dimensional imaging, we apply the method to a 3D in vitro experiment \cite{bureau2023three}. A matrix transducer array images a tissue-mimicking phantom placed behind a layer of pork tissue composed of fat and muscle (Fig.~\ref{fig:3d_invitro}). While the conventional ultrasound image fails to distinguish these layers (Fig.~\ref{fig:3d_invitro}b), the overlying tissue acts as a strong aberrator. This effect is revealed by the reflection point spread functions (RPSFs) extracted from the antidiagonals of $\mathbf{R}_{\mathbf{xx}}(z)$ (Fig.~\ref{fig:3d_invitro}c), which display a distorted confocal peak superimposed on a diffuse background arising from multiple scattering. These distortions are again captured by the focusing metric $\mathcal{F}(c)$.

We apply the DMI optimization to the corresponding open-access dataset~\cite{bureau2023three_dataset}. To accommodate the large data volume, we factorize the forward model and batch frequency components during gradient computation (Methods, Supplementary Section~\ref{sup:fact}). The converged 3D sound-speed map (Fig.~\ref{fig:3d_invitro}d) clearly discriminates fat ($c\sim1520$ m·s$^{-1}$) from muscle ($c\sim1560$ m·s$^{-1}$). It perfectly matches with the distribution of muscle and fat in the pork tissue layer. Horizontal slices (Fig.~\ref{fig:3d_invitro}e) highlight the lateral resolution and the uniformity of the recovered velocity beyond $z=15$ mm.

Correction with the optimized sound-speed model restores a diffraction-limited confocal focus (Fig.~\ref{fig:3d_invitro}c), with improved contrast relative to the diffuse background and a focusing quality increase by 49\% (see Supplementary Figure~\ref{fig:sup_fig_3_extended}). Residual multiple scattering persists, however, underscoring the limits of the current forward model. While the multilayer network captures forward multiple-scattering effects associated with long-range sound-speed fluctuations, it does not account for more complex paths induced by short-scale heterogeneities. Addressing these effects would require more advanced models, such as Born-series approaches \cite{wasik2026}, at the cost of increased sensitivity to convergence and local minima.

\vspace{5mm} 
\noindent {\large \textbf{Towards \textit{In Vivo} Applications}} 

\begin{figure*}[ht]
	\centering
	\includegraphics[width=\textwidth]{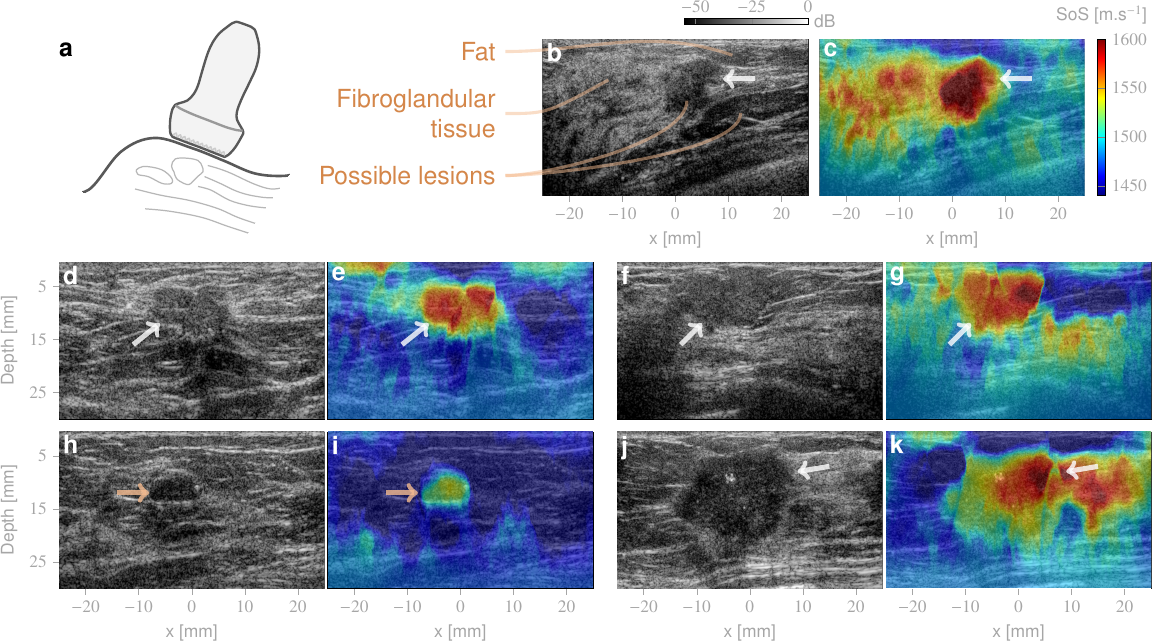}
	\caption{\label{fig:2d_invivo} \textbf{Sound speed imaging applied to breast ultrasound.}
	\textbf{a}, Schematic drawing of the acquisition, in which an ultrasound probe is positioned on top of breast tissues. \textbf{b}, \textbf{d}, \textbf{f}, \textbf{h}, \textbf{j}, Example of reflection images obtained after correction (see Supplementary Videos 3-7 for comparison between initial and corrected images). \textbf{c}, \textbf{e}, \textbf{g}, \textbf{i}, \textbf{k}, Estimated sound speed maps overlaid on the associated reflection images. Speed of sound maps are color-coded from 1420~m.s$^{-1}$ (blue) to 1600~m.s$^{-1}$ (red). White arrows indicate on each image invasive carcinoma (malignant) while orange arrows correspond to benign masses. 
	}
\end{figure*}

Ultrasound imaging is a noninvasive, portable, real-time, and cost-effective modality widely used in clinical practice, including detection and characterization of breast lesions. However, conventional B-mode images provide primarily qualitative information and require expert interpretation. Although quantitative ultrasound metrics such as shear-wave velocity have been introduced \cite{bercoff2003vivo}, their clinical adoption remains limited by hardware complexity, resolution constraints, and operator dependence.

In this context, the longitudinal wave speed is a promising parameter for tissue characterization like breast lesions assessement  \cite{schweizer2025pulse,li2009vivo}. We performed a retrospective analysis of anonymized data acquired during routine breast ultrasound examinations at Institut Bergonié (Bordeaux, France). Representative B-mode images and the corresponding speed-of-sound maps obtained with DMI are shown in Fig.~\ref{fig:2d_invivo}. In all cases, the diagonal feature of the focused reflection matrix is improved, together with an increase of the focusing quality from 27 to 53\% (see Supplementary Figure~\ref{fig:sup_fig_4_extended}). The recovered wave-speed distributions closely match anatomical features and enable clear differentiation between tissue types. Fat exhibits low sound speed ($\sim1450$ m·s$^{-1}$), benign mass and fibroglandular tissue intermediate values ($\sim1525$ m·s$^{-1}$ and $\sim1550$ m·s$^{-1}$, respectively), and malignant masses markedly higher speeds ($\sim1585$ m·s$^{-1}$; white arrows). Importantly, sound-speed contrast improves lesion discrimination beyond B-mode imaging alone, as illustrated by anechoic regions that are indistinguishable in B-mode (Fig.~\ref{fig:2d_invivo}b) but separable based on their wave speed (Fig.~\ref{fig:2d_invivo}c).

Furthermore, by compensating for aberrations, the resulting speed-of-sound maps reveal bright point-like scatterers (Supplementary Video 7) that correspond to micro-calcifications in the reflection image (Fig.~\ref{fig:2d_invivo}j). As precursors to breast cancer, these structures are vital for clinical diagnosis; however, they often remain undetected in conventional ultrasound due to aberrations induced by wave speed variations.

These findings provide a promising preliminary validation of DMI under clinical conditions for breast lesion detection and assessment. Prospective studies are required to establish sound speed as a reliable biomarker for tumor characterization. If confirmed, this approach could support ultrasound-assisted breast cancer diagnosis by reducing unnecessary biopsies and false-negative findings.

More broadly, sound speed is a versatile biomarker relevant to numerous applications, including liver disease assessment \cite{imbault2017robust, stahli2023first} and nondestructive testing \cite{Lan2014}. Because it reflects tissue composition, mechanical properties, and intensive parameters such as temperature or pressure, sound-speed imaging offers a general quantitative probe for monitoring biological and physical systems.

\vspace{5mm} 
\noindent {\large \textbf{Discussion}} 

This work demonstrates that learning-based formulations of wave propagation provide a powerful and general framework for quantitative imaging in complex media. By casting wave imaging as an optimization problem guided by a physically meaningful focusing quality metric, and by implementing the forward propagation as a multi-layer differentiable model, we bridge wave physics and modern optimization strategies in a unified framework. Rather than replacing physical modeling, our approach embeds it at the core of the learning process, enabling quantitative imaging in reflection geometries that have long been considered ill-posed.

Our approach integrates the split-step Fourier method~\cite{hardin1973applications, stoffa_splitstep_1990} into a matrix imaging framework~\cite{lambert2020reflection}. Unlike previous works that failed to properly account for lateral variations of the speed-of-sound~\cite{Bureau2024}, our approach consists in optimizing 2D and 3D wave velocity fields via end-to-end gradient ascent across the entire imaging pipeline. The main strength of this formulation lies in its scalability, optimizing from 20,000 to over 300,000 parameters, and yielding three key advances: (\textit{i}) accurate 2D speed-of-sound estimation, outperforming state-of-the-art techniques in a field actively pursued over the past decade (Supplementary Section~\ref{sec:sota}); (\textit{ii}) 3D speed-of-sound mapping in ultrasound imaging, which, to our knowledge, has not been previously achieved; and (\textit{iii}) robust tumor identification in breast ultrasound, highlighting the potential of quantitative ultrasound for cancer screening, diagnosis, and monitoring.

From an inverse-problem perspective, the proposed strategy shares conceptual similarities with full-waveform inversion in seismology~\cite{virieux2009overview} and learning-based reconstruction techniques developed in optics~\cite{kamilov2016optical,kamilov2015learning}. These approaches rely on accurate forward models and data-fidelity losses defined in the measurement space, often requiring detailed prior knowledge of the medium, sources, or initial velocity models. By contrast, our method is driven by a loss function derived from the quality of numerical focusing itself. This choice relaxes the need for detailed short-scale knowledge of the medium and enables velocity estimation directly from reflection data.

The forward model considered here relies on a multi-layer phase-screen representation that captures forward multiple scattering induced by large-scale velocity variations. 
The method presented in this work has been developed and tuned for tissues with speed-of-sound contrast ranging from 30 to 150 m/s (wave velocity variations of 2-10\%). In this limited variation range, the split-step Fourier model can be considered mostly valid to simulate wave propagation, and we are able to recover an accurate wave velocity map.

With regards to spatial resolution, diffraction theory states that the wave velocity landscape can only be retrieved over a limited part of the spatial frequency spectrum under a single back-scattering approximation~\cite{wasik2026} (Methods). This bounded support implies a theoretical lateral resolution, $\delta x \sim \lambda/(2N \! A)$, equivalent to the reflectivity image and an axial resolution governed by the depth-of-field, $\delta z \sim 2 \lambda/(N \! A^2)$ (with $N\!A=\sin\alpha$, the numerical aperture of the imaging system and $\alpha$, the maximum observation angle). This anisotropic resolution is characteristic of diffraction tomography but can be, in practice, partially mitigated by an adapted regularization strategy.

For transcranial ultrasound or seismic imaging, we do not expect our propagation model to be valid anymore since wave velocity variations are increased and reverberations can no longer be ignored. Nevertheless, a multilayered visco-elastic model~\cite{Estrada2018} could be used to grasp the elastic nature of waves in the skull~\cite{White2006} or the earth, in particular to model the longitudinal and shear components as well as their potential conversion induced by scattering and reverberation phenomena. Such a sophisticated forward model would be at the cost of a greater computational burden but can be crucial to guarantee the convergence of the optimization scheme towards a physically-grounded solution.

In optical microscopy, the single back-scattering approximation remains valid as long as the penetration depth remains below the transport mean free path, over which waves retain a directional memory. Beyond this regime, wave propagation becomes diffusive and involves complex scattering trajectories. Extending the present framework to such a regime remains challenging, but recent developments based on multi-layer Born approximations and modified Born series \cite{chen2020multi,Osnabrugge2016} suggest promising routes toward tractable models of diffuse multiple scattering.

DMI provides a natural level of interpretability that is often lacking in purely data-driven methods. It allows us to identify the fundamental limitations of the reconstruction process and to devise targeted mitigation strategies, such as confocal filtering (Fig.~\ref{fig:sup_confocal}) or regularization schemes (Methods) that stabilize convergence. At the same time, the differentiable formulation of wave propagation enables the use of modern computational tools developed for deep learning, including automatic differentiation, GPU acceleration, and scalable optimization algorithms.

The range of potential applications for DMI extends well beyond ultrasound. In optical microscopy, reflection matrix approaches have already demonstrated their ability to image biological tissues in depth~\cite{Yoon2020,najar_non-invasive_2023}. Because speckle is ubiquitous in such media, DMI could retrieve the refractive index distribution in epi-detection. In seismology, DMI could enable high-resolution velocity tomography of scattering environments such as fault zones or volcanic edifices~\cite{giraudat2024matrix}, complementing surface-wave methods and providing access to both compressional and shear wave velocities at high resolution.

Several challenges remain. Accounting for the vectorial nature of electromagnetic or elastic waves will increase computational complexity but will also unlock access to additional physical observables, such as birefringence~\cite{Villiger2016} or elastic anisotropy~\cite{Ranganathan2008}. More broadly, wave velocity is only one among other parameters governing wave propagation. The flexibility of the proposed framework makes it naturally extendable to the estimation of other material properties, including density and compressibility in acoustics or permittivity and permeability in electromagnetism.

Altogether, this work establishes learning-based matrix imaging as a unifying paradigm that combines physical insight, computational efficiency, and experimental accessibility. By exploiting reflection matrices and intrinsic focusing metrics, it opens new avenues for quantitative wave imaging across scales and modalities, in regimes that have long resisted conventional inversion strategies.

\vspace{5mm}
\noindent {\large \textbf{Methods}}

\vspace{2mm}
\noindent \textbf{Acquisition of the Reflection Matrix}

To perform an ultrasound acquisition, the ultrasonic probe emits a set of incident waves ($\mathbf{i}$) in the medium of interest. For each emission, the time-dependent back-scattered wave-field $\mathcal{R}(\mathbf{i}_{\textrm{in}}, \mathbf{o}_{\textrm{out}}, t)$ is recorded in a reception basis ($\mathbf{o}$). All recorded wave-fields are stored in a reflection matrix $\mathbf{R}_{\mathbf{i}\mathbf{o}}(t)=[\mathcal{R}(\mathbf{i}_{\textrm{in}}, \mathbf{o}_{\textrm{out}}, t)]$.

The simplest acquisition sequence is to emit with one element at a time and for each emission record with all elements the time-dependent field reflected back from the medium.  In this case, the reflection matrix $\mathbf{R}_{\mathbf{u}\mathbf{u}}(t)$ is therefore recorded in the transducer basis ($\mathbf{u}$) and its elements $\mathcal{R}(\mathbf{u}_{\textrm{in}}, \mathbf{u}_{\textrm{out}}, t)$ correspond to the impulse response between each transducer. This canonical basis is the one used in the numerical simulation described further and in the in vitro experiments~\cite{ali2023sound, bureau2023three_dataset} whose ultrasound data set has been used for the first experimental proof-of-concept described in Figs.~\ref{fig:2d_invitro} and~\ref{fig:3d_invitro}. Alternately, the reflection matrix can be acquired using beam forming (emitting and/or receiving with all elements in concert with appropriate time delays applied to each element) to form, for example, focused beams as in the conventional B-mode or plane waves of incident angle $\alpha$ for high frame rate imaging~\cite{montaldo_coherent_2009}.  This plane wave basis ($\bm{\alpha}$) was actually used as an emission basis ($\mathbf{i}=\bm{\alpha}$) to record the reflection matrix $\mathbf{R}_{\bm{\alpha}\mathbf{u}}(t)$ for the in-vivo breast data (Fig.~\ref{fig:2d_invivo}).

The acquisition parameters used in the three experiments (Figs.~\ref{fig:2d_invitro},~\ref{fig:3d_invitro},~\ref{fig:2d_invivo}) as well as in the simulation validation (Fig.~\ref{fig:sup_simu}) are reported in Tab.~\ref{tab:parameters}. Due to the need for application-specific ultrasound hardware, probes and emission parameters vary between the experiments. The 2D \textit{in vitro} acquisition was performed with a Verasonics Vantage 256 scanner (Verasonics Inc., Kirkland, WA, USA) using a L12-5 ultrasound probe~\cite{ali2023sound}. The 3D \textit{in vitro} experiment was performed with a 2D matrix array of transducers (Vermon) using an electronic hardware developed by Supersonic Imagine (Aix-en-Provence, France)~\cite{bureau2023three}. 

\begin{table*}[ht]
\centering
	\begin{tabular}{rcccc}
		& \\
		\hline
		Parameter &  2D Simulation & 2D In vitro & 2D Breast & 3D In vitro\\
		\hline
		Central frequency [in MHz] & $5.2$ & $6$ & $5$ or $6.5$  & $3$\\
		Bandwidth [at -6~dB, in \%] & $58$ & $50$ & $60$  & $66$\\
		Number of transducers & $128$ & $256$ & $256$  & $32\times 32$\\
		Pitch [in mm] & $0.3$ & $0.195$ & $0.2$ & $0.5$ \\
		Emission basis & transducer & transducer & plane wave & transducer \\
		Number of emissions & $128$ & $128$ & $61$ & $8\times 16$ \\
		
		\hline
		
		Lateral size $w$ [in $\lambda$] & $150$ & $200$ & $250$ & $60\times 40$ \\
		Axial size $h$ [in $\lambda$] & $150$ & $175$ & $150$ & $120$\\
		Grid resolution [in $\lambda$] & $0.8$ & $0.8$ & $0.8$ & $1$\\
		Padding width & $w/4$ & $w/4$ & $w/4$ & $(w/4) \times (w/4)$\\
		
		\hline
		
		Initial velocity [m.s$^{-1}$] & $1540$ & $1480$ & $1480$ & $1540$ \\
		Factorization & Yes & Yes & Yes & Yes \\
		Focusing Quality formula & equation~\eqref{eq:focusing_quality} & equation~\eqref{eq:focusing_quality} & equation~\eqref{eq:focusing_quality} & equation~(S6) \\
		Confocal filter width $w$ [in $\lambda$] & $32$ & $32$ & $32$ & $-$ \\
		
		\hline
		
		Learning rate for $\sigma_m$  & $2\cdot 10^{-3}$ & $2\cdot 10^{-3}$ & $2\cdot 10^{-3}$ & $1\cdot 10^{-3}$\\
		Learning rate for $\delta \sigma$ & $6\cdot 10^{-3}$ & $5\cdot 10^{-3}$ & $8\cdot 10^{-3}$ & $4\cdot 10^{-3}$\\
		Regularization weight ($\sigma_m$)  & $2$ & $2$ & $2$ & $1$\\
		Regularization weight ($\delta \sigma$) & $5$ & $5$ & $5$ & $20$\\
		Median filter size [in $\lambda$] & $12 \times 7$ & $12 \times 7$ & $12 \times 7$ & $5 \times 5\times 5$\\

		\hline& 
	\end{tabular}
\caption{\label{tab:parameters} \textbf{Experimental parameters.}
		Acquisition parameters, spatial grid parameters, algorithm choices and optimization parameters for the four different experiments. 2D simulations correspond to Fig.~\ref{fig:sup_simu}, 2D in vitro experiment to Fig.~\ref{fig:2d_invitro}, 2D breast experiment to Fig.~\ref{fig:2d_invivo} and 3D in vitro experiment to Fig.~\ref{fig:3d_invitro}. }
\end{table*} 

For breast tissue data, we have complied with all relevant ethical regulations following standard practice defined by General Data Protection Regulation (GDPR) and Commission nationale de l’information et des libertés (CNIL). This retrospective study analyzed existing, fully anonymized ultrasound data collected during routine care on patients from which informed consent had been obtained. The acquisition was performed using a medical ultrafast ultrasound scanner (Aixplorer Mach-30, Supersonic Imagine, Aix-en-Provence, France). The skin was placed in direct contact of a linear array of transducers (SL18-3, Supersonic Imagine) whose characteristics are also provided in Tab.~\ref{tab:parameters}.

\vspace{2mm}
\noindent \textbf{Numerical Focusing}

To numerically focus the acquired signals at a given point in the medium, usual imaging techniques model the propagation of emitted and received waves and apply their time-reversed counterparts to the reflection matrix $\mathbf{R}_{\mathbf{i o}}(t)$.
Such an operation is denoted as time-reversal, back-projection, adjoint imaging or migration in the different fields of wave physics.
In this paper, this focusing operation is performed in the frequency domain by considering the transmission matrix $\mathbf{T}_{\mathbf{r i}}(f; c)=[\mathcal{T}(\mathbf{r}_{\textrm{in}}, \mathbf{i}_{\textrm{in}}, f; c)]$ that links, at each frequency $f$ and according to the velocity model $c$, the emission basis ($\mathbf{i}$) to the focused basis ($\mathbf{r}$) that contains a set of focusing points $\mathbf{r}_{\textrm{in}}$ inside the medium under study. Each column of the transmission matrix corresponds to the incident wave-field at each point $\mathbf{r}_{\textrm{in}}$ induced by each emitted wave $\mathbf{i}_{\textrm{in}}$ for a wave velocity model $c$.

This transmission matrix depends on the chosen propagation model and is of the upmost importance for the numerical focusing process. Its expression in the case of an advanced propagation model (split-step Fourier method) will be discussed in details in the next section. In the simple case of a homogeneous velocity model $c(\mathbf{r}) = c_0$ for all $\mathbf{r}$, its expression can be found analytically from the homogeneous wave equation. For instance, for a point-like emission at position $\mathbf{u}_{\textrm{in}}$, we obtain the Green's functions of the wave equation in 2D:
\begin{equation}
    \mathcal{T}(\mathbf{r}_{\textrm{in}}, \mathbf{u}_{\textrm{in}}, f; c_0) = 
    -\frac{j}{4}\mathcal{H}_0^{(1)}\left(k_0 |\mathbf{u}_{\textrm{in}} - \mathbf{r}_{\textrm{in}}|\right),
\end{equation}
with $k_0=2\pi f/c_0$ the wavenumber and $\mathcal{H}_0^{(1)}(-)$ the Hankel function of the first kind. Similarly, for a plane wave emission of propagation angle $\alpha_{\textrm{in}}$ in 2D, we have:
\begin{equation}
    \mathcal{T}(\mathbf{r}_{\textrm{in}}, \alpha_{\textrm{in}}, f; c_0) = 
    \exp{\left [j k_0 \begin{pmatrix}
    \sin\alpha_{\textrm{in}} \\
    \cos\alpha_{\textrm{in}}
    \end{pmatrix} \cdot \mathbf{r}_{in}\right]},
\end{equation}
with $\cdot$ the scalar product and $j=\sqrt{-1}$ the unitary imaginary number.
As per the emission case, we can define a reception transmission matrix $\mathbf{T}_{\mathbf{r}\mathbf{o}}(f; c)=[\mathcal{T}(\mathbf{r}_{\textrm{out}}, \mathbf{o}_{\textrm{out}}, f; c)]$, between the reception basis ($\mathbf{o}$) and the focused basis ($\mathbf{r}$).

To perform the numerical focusing, we multiply the time-reversed counterpart of the transmission responses $\mathcal{T}$ (i.e. their complex conjugates in the Fourier domain) with each frequency component $\mathcal{R}(\mathbf{i}_{\textrm{in}}, \mathbf{o}_{\textrm{out}}, f)$ of the acquired reflection matrix. A sum over all emissions and receptions allows us to perform transmit and receive focusing operations in post-processing. A sum over frequency amounts to time gate the backscattered signal at an echo time fixed by the wave velocity model.
We finally obtain a reflection (confocal) image $\mathcal{I}(\mathbf{r}; c)$ for any position $\mathbf{r}$:
\begin{equation}
	\mathcal{I}(\mathbf{r}; c) = \sum\limits_f \sum\limits_{\substack{\mathbf{i}_{\textrm{in}}\in\mathbf{i}\\ \mathbf{o}_{\textrm{out}}\in\mathbf{o}}} \mathcal{T}^*(\mathbf{r}, \mathbf{i}_{in}, f; c) \mathcal{R}(\mathbf{i}_{in}, \mathbf{o}_{out}, f) \mathcal{T}^*(\mathbf{r}, \mathbf{o}_{out}, f; c).
	\label{eq:numerical_focusing_confocal2} 
\end{equation}
where the symbol $*$ stands for phase conjugate.

The matrix imaging framework goes beyond traditional confocal imaging as it separates the focusing points in emission and reception. In practice, it consists in considering different focal spots $\mathbf{r}_{in}$ and $\mathbf{r}_{out}$ in the transmission functions $\mathcal{T}$:
\begin{equation}
	\overline{\mathcal{R}}(\mathbf{r}_{in}, \mathbf{r}_{out}; c) = \sum\limits_f \sum\limits_{\substack{\mathbf{i}_{in}\in\mathbf{i}\\ \mathbf{o}_{out}\in\mathbf{o}}} \mathcal{T}^*(\mathbf{r}_{in}, \mathbf{i}_{in}, f; c) \mathcal{R}(\mathbf{i}_{in}, \mathbf{o}_{out}, f) \mathcal{T}^*(\mathbf{r}_{out}, \mathbf{o}_{out}, f; c),
	\label{eq:numerical_focusing}
\end{equation}
or, under a matrix formalism,
\begin{equation}
\label{foc2}
    \overline{\mathbf{R}}_{\mathbf{rr}}(c)=\sum_f \mathbf{T}_{\mathbf{ri}}^*(f;c) \times \mathbf{R}_{\mathbf{io}}(f)  \times  \mathbf{T}_{\mathbf{ro}}^{\dag}(f;c).
\end{equation}
with $\dagger$ the hermitian transpose operation and $\times$ the matrix product.
\noindent
The time-gated focused reflection matrix $ \overline{\mathbf{R}}_{\mathbf{rr}}=[	\overline{\mathcal{R}}(\mathbf{r}_{\textrm{in}}, \mathbf{r}_{\textrm{out}}; c) ]$ contains the cross-talks between the emission and reception focusing points that act as virtual sources $\mathbf{r}_{\textrm{in}}$ and receivers $\mathbf{r}_{\textrm{out}}$. When the emission and reception focal points coincide ($\mathbf{r}_{in}= \mathbf{r}_{out}$), we fall back to the confocal image $\mathcal{I}$. The confocal image actually corresponds to the diagonal coefficients of $ \overline{\mathbf{R}}_{\mathbf{rr}}$:
\begin{equation}
	\mathcal{I}(\mathbf{r}; c) = \overline{\mathcal{R}}(\mathbf{r},\mathbf{r}; c) = \text{diag}\left(\overline{\mathbf{R}}_{\mathbf{rr}}(c)\right).
	\label{eq:numerical_focusing_confocal} 
\end{equation}

In practice,  we can restrict our study to pairs of virtual transducers,  $\mathbf{r}_{\textrm{in}}=(\mathbf{x}_{\textrm{in}},z)$
and  $\mathbf{r}_{\textrm{out}}=(\mathbf{x}_{\textrm{out}},z)$, located at the same depth $z$. A focused sub-matrix $\overline{\mathbf{R}}_{\mathbf{xx}}(z;c)=[\overline{\mathcal{R}}(\mathbf{x}_{\textrm{in}},\mathbf{x}_{\textrm{out}},z;c)]$ can then be considered at each depth $z$:
\begin{equation}
	\overline{\mathbf{R}}_{\mathbf{xx}}(z; c) = \sum\limits_f \mathbf{T}_{\mathbf{x}\mathbf{i}}^*(z, f; c) \times \mathbf{R}_{\mathbf{i}\mathbf{o}}(f) \times \mathbf{T}_{\mathbf{x}\mathbf{o}}^\dagger(z, f; c),
	\label{eq:numerical_focusing_matrix}
\end{equation}
Similarly to the full focused reflection matrix $\overline{\mathbf{R}}_{\mathbf{rr}}(c)$ (Eq.~\ref{eq:numerical_focusing_confocal}), the diagonal of each sub-matrix $	\overline{\mathbf{R}}_{\mathbf{xx}}(z; c) $ corresponds to the confocal reflection image at depth $z$.

\vspace{2mm}
\noindent \textbf{Forward Model}

The modeling of the transmission matrices, $\mathbf{T}_{\mathbf{r}\mathbf{i}}(f; c)$ and $\mathbf{T}_{\mathbf{r}\mathbf{o}}(f; c)$, is crucial to reach a satisfying focusing process in heterogeneous media.
On one hand, it needs to be sufficiently simple so that the optimization process is tractable. On the other hand, it needs to be accurate enough to focus efficiently in complex media.
State of the art solutions use drastic assumptions on wave propagation to simplify the transmission matrix. For instance, wave propagation can be modeled as a straight way propagation \cite{simson2023differentiable, simson2025ultrasound} or as a propagation in a homogeneous medium with one or several phase screens \cite{lambert2020reflection, bureau2023three, heriard2025path}.
In our case, we want to address highly heterogeneous media and therefore choose a propagation model that takes into account advanced wave propagation phenomena such as refraction and forward multiple scattering. To do so, we use the split-step Fourier method \cite{hardin1973applications}, also called the beam propagation method in optics \cite{feit1978light}. This method models wave propagation as a succession of phase screens and free-field propagation at the wavelength level, and approximates accurately the wave equation solution when the wave propagates in a preferred direction.

The implementation of the split-step Fourier method starts with an interpolation whose goal is to express the experimental reflection matrices $\mathbf{R}_{\mathbf{i}\mathbf{o}}(f)$ as a set of focused reflection matrices $\mathbf{R}_{\mathbf{x}\mathbf{x}}(z_0, f)$ on a starting plane $z=0$ chosen to be the sensor plane. We use either a spatial interpolation if emissions or receptions are performed in the transducer basis, or an inverse Fourier transforms in the case of plane wave emissions.
We denote this interpolation operator as a matrix $\mathbf{I}_{\mathbf{x}\mathbf{i}}(f)$ (respectively $\mathbf{I}_{\mathbf{x}\mathbf{o}}(f)$ for the reception side), leading to the following formula:
\begin{equation}
    \mathbf{R}_{\mathbf{x}\mathbf{x}}(0, f; c) = \mathbf{I}_{\mathbf{x}\mathbf{i}}(f; c) \times \mathbf{R}_{\mathbf{i}\mathbf{o}}(f) \times \mathbf{I}^T_{\mathbf{x}\mathbf{o}}(f; c).
    \label{eq:split_step_init}
\end{equation}

With the obtained focused reflection matrices $\mathbf{R}_{\mathbf{x}\mathbf{x}}(0, f)$ at plane $z=0$ and for each frequency $f$, we decompose the transmission function as a series of elementary operators that model the propagation in a slice of height $\delta z$. These elementary operators depend on the wave velocity model at depth $z$ denoted as $c(z)$ and are written $\mathbf{\bm{\delta} T}_{\mathbf{x}\mathbf{x}}\left(\delta z, f; c(z)\right)$. We obtain for the propagation between planes $z$ and $z+\delta z$:
\begin{equation}
    \mathbf{R}(z + \delta z, f; c) =
    \mathbf{\bm{\delta} T}^*\left(\delta z, f; c(z)\right) \times \mathbf{R}(z, f; c) \times
    \mathbf{\bm{\delta} T}^\dagger\left(\delta z, f; c(z)\right),
	\label{eq:split_step_iteration}
\end{equation}
where all subscripts $-_{\mathbf{x}\mathbf{x}}$ have been removed for the sake of simplicity.
The elementary operator $\mathbf{\bm{\delta} T}$ is defined as a free-space propagation performed in the frequency domain followed by a phase screen modeling the lateral variations of the wave velocity:
\begin{equation}
\label{eq0}
    \mathbf{\bm{\delta} T}\left(\delta z, f; c(z)\right) = \mathbf{s}(f; c(z)) \odot \left(\mathbf{F}^\dagger \times \left( \mathbf{h}(f, c(z)) \odot \mathbf{F} \right)\right),
\end{equation}
where the symbol $\odot$ stands for the Hadamard (element wise) product. $\mathbf{F}=[F(\mathbf{f_x},\mathbf{x})]$ is the spatial Fourier transform (implemented with the fast Fourier transform in practice), such that:
\begin{equation}
    F(\mathbf{f_x},\mathbf{x})=\exp \left (-j 2 \pi \mathbf{f_x} \cdot \mathbf{x} \right ),
\end{equation}
with $\mathbf{f_{x}}$ the lateral spatial frequencies, written as a vector.
$\mathbf{h}(f,c_m(z))=[h(\mathbf{f_x},f,c_m(z))]$ is the free-space propagation operator over depth $\delta z$ defined for a constant wave speed $c_m(z)$. $\mathbf{s}(f; c(z))=[s(\mathbf{x},f; c(z))]$ is the phase screen operator defined in real space. In Eq.~\ref{eq0}, all upper case bold letters represent matrices and all lower case bold letters represent vectors.
The propagation operator $\mathbf{h}$ and the phase screen $\mathbf{s}$ can be expressed as a function of the slowness, defined as the inverse of the velocity, $\sigma(\mathbf{x},z)=1/c(\mathbf{x},z)$. More precisely, $\mathbf{h}$ is dictated by the average slowness $\sigma_m(z) = \langle \sigma(\mathbf{x}, z)\rangle_{\mathbf{x}}=1/c_m(z)$ and $\mathbf{s}$ by its residual $\delta \sigma (\mathbf{x}, z) = \sigma(\mathbf{x}, z) - \sigma_m(z)$ at depth $z$, such that:
\begin{align}
    \mathbf{h}(f, c(z)) &= \exp{\left(j 2\pi \delta z \sqrt{\left(f \sigma_m(z) \right)^2 - |\mathbf{f_{x}}|^2}\right)},
    \label{eq:free_space_prop}
    \\
        \mathbf{s}(f, c(z)) &= \exp{\left(j 2\pi \delta z 
     f \delta \sigma(\mathbf{x}, z)\right)},
     \label{eq:phase_screen}
\end{align}
with $\sqrt{-}$ the complex square root.

The free-space propagation is performed in the spatial Fourier space which supposes periodic boundaries. In practice, we prefer to consider absorbing boundaries that can be implemented using a lateral padding and adding an apodization step at the end of the elementary propagation~(equation \eqref{eq:split_step_iteration}). This apodization is performed by multiplying the result of the propagation by a function equal to $1$ in the domain of interest, and decreasing to 0 continuously in the padding section (e.g. using a Tukey window).

With this recursive implicit definition of the transmission matrix, the experimental reflection matrix is first projected on plane $z=0$ with equation~\eqref{eq:split_step_init} and then propagated iteratively at all depths with equation~\eqref{eq:split_step_iteration}. At each depth, the obtained focused reflection matrices can be summed over temporal frequencies to perform time-gating and compute the focusing quality (equation~\eqref{eq:focusing_quality}).

The resulting forward model does not rely on ray tracing and takes into account advanced propagation effects such as refraction and diffraction. To assess the role of this wave propagation model in the overall optimization scheme, Supplementary Section~\ref{sec:sota} compares our forward model with the standard straight ray propagation model. Supplementary Figure~\ref{fig:sup_model} clearly shows the importance of accounting for refraction and diffraction to retrieve a high-resolution and high-contrast velocity map.

\vspace{2mm}
\noindent \textbf{Focusing Quality}

The focused reflection matrix provides a self-portrait of the focusing process and have been studied for multiple applications such as ultrasound imaging \cite{lambert2020reflection, bureau2023three}, optical coherence microscopy \cite{balondrade2024multi} or passive seismic imaging \cite{giraudat2024matrix}.
In particular, it can be used to assess the numerical focusing efficiency as the width of its diagonal is representative of the effective size of the focal spots. The profile along the antidiagonal of the focused reflection matrices is called the reflection point-spread-function or RPSF \cite{lambert2020reflection}. More specifically, echoes that have been successfully focused lie on the diagonal of the focused reflection matrix whereas echoes resulting from aberrations or multiple scattering, that are not adequately grasped by the forward model and/or not correctly inverted by the focusing process, spread outside of diagonal. We devise a focusing quality criterion by averaging the energy on the diagonal of the focused reflection matrix, normalized by the energy averaged over each column of the focused reflection matrix:
\begin{equation}
    \mathcal{F}(c) = \left\langle \frac{|\overline{\mathcal{R}}(\mathbf{x}, \mathbf{x}, z; c)|^2}{\sum\limits_{\mathbf{x'}} |\overline{\mathcal{R}}(\mathbf{x}', \mathbf{x}, z; c)|^2} \right\rangle_{\mathbf{x}, z},
    \label{eq:focusing_quality}
\end{equation}
where $\mathbf{x}$ represents the lateral coordinates [$x$ in 2D and $(x, y)$ in 3D].
This criterion is representative of the diagonal feature of the focused reflection matrix and therefore of the efficiency of the numerical focusing process. With this matrix imaging framework, numerical focusing and the quality factor can be used equivalently in 2D or 3D configurations. In the latter case, the position $\mathbf{r}$ corresponds to three coordinates $(x, y, z)$ and the focused reflection matrices are four-dimensional.

In the proposed approach, the focusing quality criterion can be switched with other state-of-the-art metrics. We study such alternatives in Supplementary Section~\ref{sec:sota} and show that the focusing quality yields the best results with the considered forward model  (Supplementary Figure \ref{fig:sup_loss}).
Additionally, the use of such a matrix imaging derived criterion enables the use of a confocal filter to further improve the results in clinical conditions.

\vspace{2mm}
\noindent \textbf{Confocal Filter}

To improve the robustness of this focusing quality criterion, especially for large domains or \textit{in vivo} configurations, we apply a confocal filter to the focused reflection matrix before computing the quality criterion. 
This confocal filter decreases the contribution of random multiple scattering trajectories that give rise to a drastic off-diagonal spreading of the RPSF and that can not be grasped by our forward model.
 For a confocal filter of width $w$ (typically a few tens of the wavelength $\lambda$), we obtain a filtered focused reflection matrix $\overline{\mathbf{R}}_{\mathbf{x}\mathbf{x}}(z; c)$ defined as:
\begin{equation}
\label{confocal}
    \overline{\mathcal{R}}(\mathbf{x}_{in}, \mathbf{x}_{out}, z; c) =
    \begin{cases}
        \overline{\mathcal{R}}(\mathbf{x}_{in}, \mathbf{x}_{out}, z; c) &\text{if}\quad |x_{in} - x_{out}| < w,\\
        0 &\text{else}.
    \end{cases} 
\end{equation}
The values $w$ used for the confocal filter are reported for each experimental or numerical configurations in Tab.~\ref{tab:parameters}.

This filtering is crucial for the applications of differentiable beamforming methods to in-vivo data sets like the breast data used in the current paper. In its absence, diffuse multiple scattering and clutter can prevent the  convergence of our optimization scheme. Figure~\ref{fig:sup_confocal} illustrates the last statement by comparing the wave velocity map estimated with (Fig.~\ref{fig:sup_confocal}c) and without a confocal filter (Fig.~\ref{fig:sup_confocal}b). In the latter case, the obtained velocity map fails to converge, most likely due to strong afocal signal (e.g. multiple scattering, reverberation, noise) that cannot be grasped by the split-step model and therefore hinders its convergence.\\
\begin{figure*}[htp]
	\includegraphics[width=\textwidth]{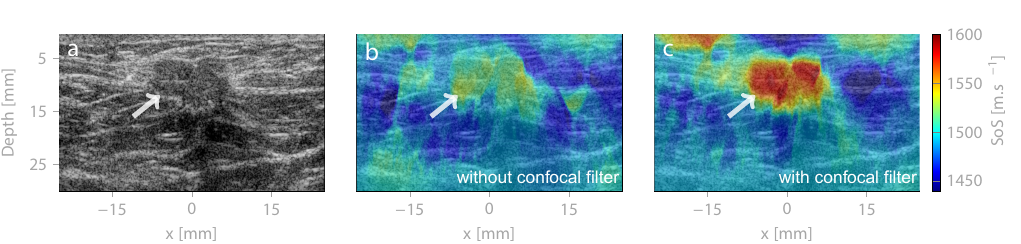}
	\caption{\textbf{Impact of the confocal filter}. The role of the confocal filter is illustrated by considering one example of a breast tissue data set (Tab.~\ref{tab:parameters}). \textbf{a}, Reflectivity image. \textbf{b}, Overlaid wave velocity estimated without any confocal filter. \textbf{c}: Overlaid wave velocity estimated with a confocal filter applied to the focused reflection matrix (Eq.~\ref{confocal}, $w=32 \lambda$).}\label{fig:sup_confocal}
\end{figure*}

\vspace{2mm}

\noindent \textbf{Optimization and Convergence}

In the previous paragraphs, we defined the operations applied to the raw data, given a velocity model, to obtain a focusing quality criterion. This forward model is represented in the upper row of Figure~\ref{fig:method}.

Our methods optimizes this focusing quality, solving the following optimization problem:
\begin{equation}
    	\max\limits_{\sigma} \mathcal{F}(\sigma),
	\label{eq:opt_pb_naive}
\end{equation}
with $\sigma$ the slowness map. We choose to solve this problem using a gradient ascent algorithm which iteration step consists in updating the slowness as $\sigma\leftarrow\sigma + \epsilon \nabla_\sigma \mathcal{F}(\sigma)$, with $\epsilon$ a learning rate and $\nabla_\sigma$ the gradient operator with respect to $\sigma$ (implemented with an Adam optimizer in practice).

To apply this algorithm, our first step is to compute $\nabla_\sigma \mathcal{F}(\sigma)$. We do so with the chain rule as in equation~\eqref{eq:gradient}, using the auto-differentiable capacities of Pytorch, inspired by recent works in adaptive ultrasound imaging \cite{simson2023differentiable, spainhour2024optimization, heriard2025path, simson2025ultrasound}. Nevertheless, the forward model described above is not differentiable everywhere and the chain rule is not tractable in practice. To solve these issues, we slightly modify the forward model and adapt the way gradients are computed (see Supplementary Section~\ref{sup:diff}).

The obtained optimization problem (equation \eqref{eq:opt_pb_naive}) is non-linear and likely ill-posed. There is no theoretical guarantees on the convergence of the gradient ascent algorithm nor on the fact that the global maximum of the focusing quality factor is associated with the actual slowness map. Such a guarantee exists for homogeneous media \cite{garnier2025probing} but has not been demonstrated for heterogeneous media yet.
Nevertheless, we can study the convergence of the gradient ascent algorithm from an empirical standpoint.
In particular, we observe convergence issues due to local minima containing short-scale fluctuations of the wave velocity map. To avoid these minima, we use a regularization as well as a filtering scheme along the iterations of the gradient ascent (see Supplementary Section~\ref{sup:tv_filt}).

If the regularization and the filtering solve most convergence issues, an asymmetry of the forward problem remains.
Indeed, the focusing quality is affected by phase aberrations accumulated during wave propagation. This accumulation corresponds to axial integrals of the slowness map, resulting in difficulties to recover the high-frequency axial variations of the wave velocity. This limitation is linked to the so-called missing cone problem in diffraction tomography \cite{Tam1981}. In our case, the computation of the focusing quality at all depths and the presence of scattering and refraction effects mitigate the missing cone problem and enables us to accurately recover an axial profile.
However, this asymmetry still leads to a slow axial convergence compared to the estimation of lateral variations.
To solve this limitation, we perform a relaxation of the optimization problem, duplicating the optimization variable into an axial profile $\sigma_m$ and a map $\delta \sigma$ of the slowness (see Supplementary Section~\ref{sup:relax} for details). Our relaxed optimization problem is of the form:
\begin{equation}
		\max\limits_{\delta \sigma, \sigma_m} \mathcal{F}(\delta \sigma, \sigma_m) - \eta TV(\delta \sigma, \sigma_m),
	\label{eq:opt_pb_final}
\end{equation}
with $TV(-)$ the regularization term introduced in Supplementary Section~\ref{sup:tv_filt}. We solve this problem in two phases, first optimizing only the axial slowness and then optimizing both simultaneously. The whole optimization pipeline has been validated by means of numerical simulations whose results are shown in Fig.~\ref{fig:sup_simu} and discussed in Supplementary Section~\ref{sec:simulation}. In each numerical configuration, the final algorithm converges towards a speed-of-sound map in good agreement with its ground truth at shallow depth ($z<25$ mm).
\begin{figure*}[ht]
	\centering
	\includegraphics[width=0.9\textwidth]{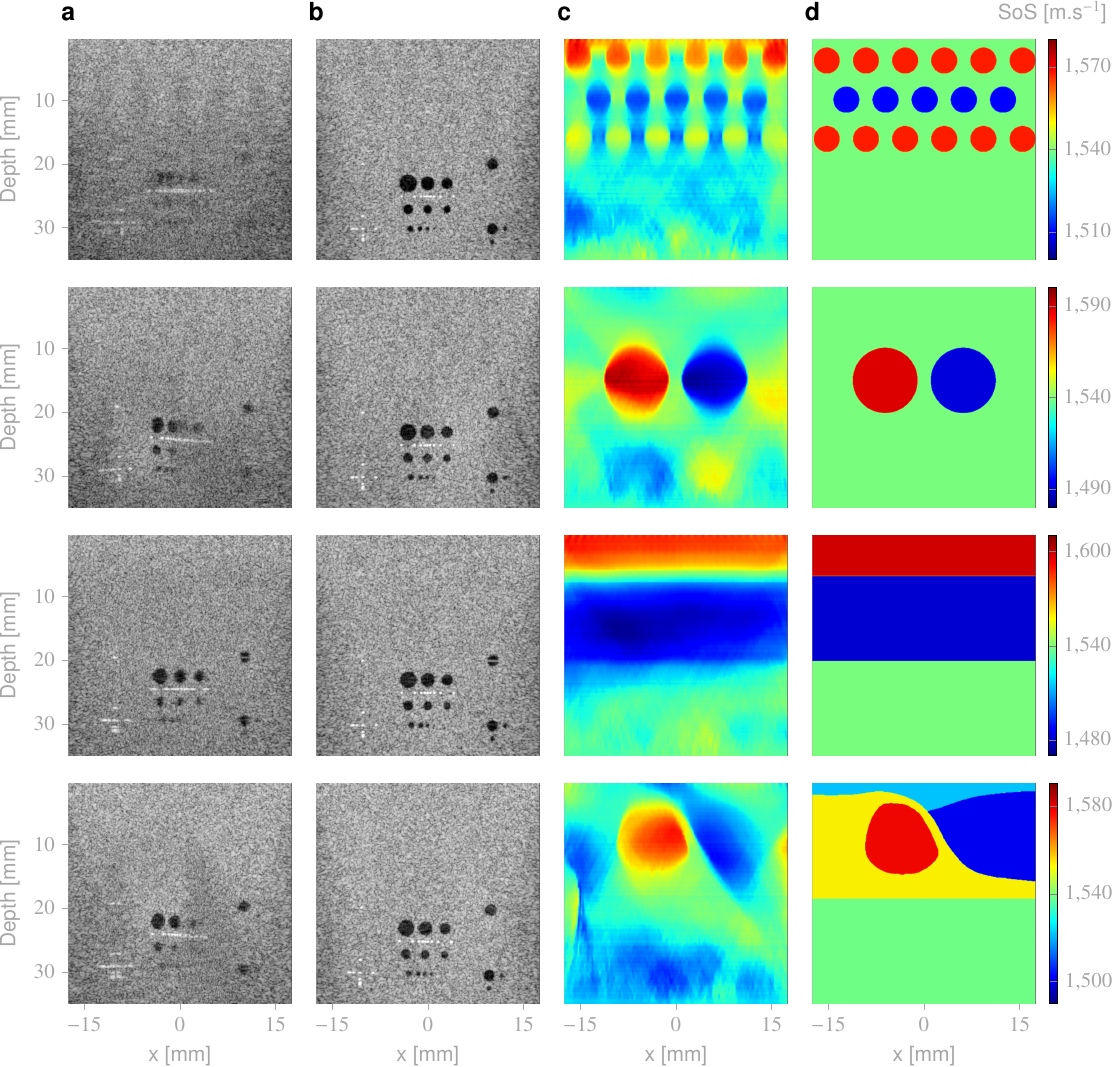}
	\caption{\label{fig:sup_simu} \textbf{Numerical validation of differential matrix imaging.}
	Four different media are simulated using k-Wave with heterogeneous velocity maps. \textbf{a}, Reflectivity images beamformed with a constant speed-of-sound ($c=1540$ m.s$^{-1}$). \textbf{b}, Reflectivity images obtained by differential matrix imaging. \textbf{c}, Corresponding speed-of-sound map. \textbf{d}, Simulated speed-of-sound landscape (ground truth). Reflection images (a,b) are represented in dB and speed of sound values (c,d) are given in~m.s$^{-1}$.}
\end{figure*}

Finally, the optimization parameters are fine-tuned empirically for each configuration (see Tab.~\ref{tab:parameters} and Supplementary Section~\ref{sec:parameters}). The number of iterations is set to $100$ for each phase of the convergence process (axial and lateral), with an early stopping criterion based on the average evolution of the loss in 10 consecutive iterations. 

In the breast tissue case, the long probe and shallow depths of imaging result in an unusual geometry with a width over height ratio of $1.7$. In this case, the decomposition of the slowness map as an axial profile plus a lateral residual slowness shows limitations. To recover a configuration similar to the other cases, we decompose our medium into three overlapping columns $125$ wavelengths wide, and perform the optimization on these three columns independently. A final optimization is performed on the full domain, with the result of the previous optimizations used as the initial velocity model.

\vspace{2mm}
\noindent \textbf{Computational Complexity and Factorization}

The calculation of the focused reflection matrix at each depth can be computationally expensive and requires a large amount of memory. This is particularly the case for the 3D configuration where the presence of two lateral dimensions leads to 4D focused reflection matrices. As the number of emissions is usually lower than the number of lateral positions in ultrasound imaging, we leverage a factorization of the reflection matrix to reduce this computational cost (see Supplementary~Section \ref{sup:fact}).
In the 3D case, we go even further and use a modified focusing quality metric that directly uses the factorized form of the reflection matrix. This modified metric is no longer strictly equivalent to the one defined in equation \eqref{eq:focusing_quality}, but also probes the focusing quality~\cite{mallart1994adaptive}.
The computational complexity and memory requirements are reported in Supplementary Section~\ref{sec:complexity} for the different configurations, together with computational tricks to reduce the memory requirements.
\\

\vspace{2mm}
\noindent \textbf{Initial velocity}

The initial velocity model is always chosen homogeneous, with a value that corresponds to the expected average value (e.g. $1480$~m.s$^{-1}$ or $1540$~m.s$^{-1}$). A variation of the initial velocity does not result in a bias (for reasonable variations, i.e lower than $10\%$) but may lead to increased convergence time and spatial instabilities at high depths, as detailed in Supplementary Section~\ref{sec:parameters}.\\

\vspace{2mm}
\noindent \textbf{Spatial sampling}

The spatial sampling of the medium used for numerical focusing and wave velocity mapping has been adapted to each experimental configuration (see Tab.~\ref{tab:parameters}). 
It is the result of a trade-off between reconstruction accuracy and computation time.
Here, isotropic grids of 0.8$\lambda$ have been used for the 2D configurations (Figs.~\ref{fig:2d_invitro}, \ref{fig:2d_invivo} and ~\ref{fig:sup_simu}) and of one wavelength for the 3D case (Fig.~\ref{fig:3d_invitro}), the considered wavelength $\lambda$ being estimated at the central frequency. However, note that this reconstruction grid could have been down-sampled as described in Supplementary Section~\ref{sec:parameters}. Indeed, as shown by diffraction theory~\cite{wasik2026}, the spatial frequency spectrum of the wave velocity landscape can be captured over a limited support under a single back-scattering approximation: $|f_x|<  NA/\lambda$ and $|f_z|< N\!A^2/(4\lambda)$. This property can be used to reduce the size of the considered grid and thus decrease the computation time. Shannon Nyquist theorem applied to the wave velocity landscape actually implies for the reconstruction grid the same lateral resolution as the reflectivity image, $\delta x \sim \lambda/(2N \! A)$, and an axial resolution governed by the depth-of-field, $\delta z \sim 2 \lambda/(N \! A^2)$. Since the numerical aperture is controlled by the transducer directivity ($\alpha \sim 35$ deg) at shallow depths, this implies a required lateral sampling, $\delta x \sim \lambda$, and an axial sampling, $\delta z \sim 6\lambda$, values that are confirmed numerically in Supplementary Figure~\ref{fig:sup_res}.\\

\vspace{2mm}
\noindent \textbf{Data Availability Statement.}
Data generated in this paper have been deposited on Zenodo at \href{https://zenodo.org/records/23419267}{https://zenodo.org/records/23419267}(ref.~\cite{heriard2026})\\

\vspace{2mm}

\noindent \textbf{Code Availability Statement.}
The corresponding code, capsule environment, and software settings used for data analysis and figure generation are openly available on Code Ocean at [Code Ocean DOI/URL when available].

\clearpage

\bibliography{my_bib2}

@Article{wasik2026,
  author  = {Wasik, T. and Barolle, V. and Aubry, A. and Garnier, J.},
  journal = {IEEE Trans. Comput. Imaging},
  title   = {Taking advantage of multiple scattering for Optical Reflection Tomography},
  year    = {2026},
  pages   = {533-547},
  volume  = {12},
}

@article{hardin1973applications,
	title={Applications of the split-step Fourier method to the numerical solution of nonlinear and variable coefficient wave equations},
	author={Hardin, Ronald H},
	journal={Siam Rev.},
	volume={15},
	pages={423},
	year={1973}
}

@Article{feit1978light,
  author    = {Feit, MD and Fleck Jr, JA},
  journal   = {Appl. Opt.},
  title     = {Light propagation in graded-index optical fibers},
  year      = {1978},
  number    = {24},
  pages     = {3990--3998},
  volume    = {17},
  publisher = {Optical Society of America},
}

@Article{rudin1992nonlinear,
  author  = {Rudin, Leonid I and Osher, Stanley and Fatemi, Emad},
  journal = {Phys. D: Nonlinear Phenom.},
  title   = {Nonlinear total variation based noise removal algorithms},
  year    = {1992},
  number  = {1-4},
  pages   = {259--268},
  volume  = {60},
}

@article{mallart1994adaptive,
	title={Adaptive focusing in scattering media through sound-speed inhomogeneities: The van Cittert Zernike approach and focusing criterion},
	author={Mallart, Raoul and Fink, Mathias},
	journal={J. Acoust. Soc. Am.},
	volume={96},
	number={6},
	pages={3721--3732},
	year={1994},
}

@Article{bercoff2003vivo,
  author    = {Bercoff, J and Chaffai, S and Tanter, M and Sandrin, L and Catheline, S and Fink, M and Gennisson, JL and Meunier, M},
  journal   = {Ultrasound Med. Biol.},
  title     = {In vivo breast tumor detection using transient elastography},
  year      = {2003},
  number    = {10},
  pages     = {1387--1396},
  volume    = {29},
  publisher = {Elsevier},
}

@article{li2009vivo,
	title={In vivo breast sound-speed imaging with ultrasound tomography},
	author={Li, Cuiping and Duric, Nebojsa and Littrup, Peter and Huang, Lianjie},
	journal={Ultrasound Med. Biol.},
	volume={35},
	number={10},
	pages={1615--1628},
	year={2009},
}

@article{virieux2009overview,
	title={An overview of full-waveform inversion in exploration geophysics},
	author={Virieux, Jean and Operto, St{\'e}phane},
	journal={Geophysics},
	volume={74},
	number={6},
	pages={WCC1--WCC26},
	year={2009},
}

@Article{treeby2010k,
  author    = {Treeby, Bradley E and Cox, Benjamin T},
  journal   = {J. Biomed. Opt.},
  title     = {k-Wave: MATLAB toolbox for the simulation and reconstruction of photoacoustic wave fields},
  year      = {2010},
  number    = {2},
  pages     = {021314--021314},
  volume    = {15},
  publisher = {Society of Photo-Optical Instrumentation Engineers},
}

@article{kamilov2015learning,
	title={Learning approach to optical tomography},
	author={Kamilov, Ulugbek S and Papadopoulos, Ioannis N and Shoreh, Morteza H and Goy, Alexandre and Vonesch, Cedric and Unser, Michael and Psaltis, Demetri},
	journal={Optica},
	volume={2},
	number={6},
	pages={517--522},
	year={2015},
	publisher={Optical Society of America}
}

@Article{kamilov2016optical,
  author  = {Kamilov, Ulugbek S and Papadopoulos, Ioannis N and Shoreh, Morteza H and Goy, Alexandre and Vonesch, Cedric and Unser, Michael and Psaltis, Demetri},
  journal = {IEEE Trans. Comput. Imag.},
  title   = {Optical tomographic image reconstruction based on beam propagation and sparse regularization},
  year    = {2016},
  number  = {1},
  pages   = {59--70},
  volume  = {2},
}

@Article{imbault2017robust,
  author  = {Imbault, Marion and Faccinetto, Alex and Osmanski, Bruno-F{\'e}lix and Tissier, Antoine and Deffieux, Thomas and Gennisson, Jean-Luc and Vilgrain, Val{\'e}rie and Tanter, Micka{\"e}l},
  journal = {Phys. Med. Biol.},
  title   = {Robust sound speed estimation for ultrasound-based hepatic steatosis assessment},
  year    = {2017},
  number  = {9},
  pages   = {3582},
  volume  = {62},
}

@Article{paszke2019pytorch,
  author  = {Paszke, Adam and Gross, Sam and Massa, Francisco and Lerer, Adam and Bradbury, James and Chanan, Gregory and Killeen, Trevor and Lin, Zeming and Gimelshein, Natalia and Antiga, Luca and others},
  journal = {Adv. Neural Inf. Process. Syst.},
  title   = {Pytorch: An imperative style, high-performance deep learning library},
  year    = {2019},
  volume  = {32},
}

@article{stahli2020improved,
	title={Improved forward model for quantitative pulse-echo speed-of-sound imaging},
	author={St{\"a}hli, Patrick and Kuriakose, Maju and Frenz, Martin and Jaeger, Michael},
	journal={Ultrasonics},
	volume={108},
	pages={106168},
	year={2020},
	publisher={Elsevier}
}

@article{lambert2020reflection,
	title={Reflection matrix approach for quantitative imaging of scattering media},
	author={Lambert, William and Cobus, Laura A and Couade, Mathieu and Fink, Mathias and Aubry, Alexandre},
	journal={Phys. Rev. X},
	volume={10},
	number={2},
	pages={021048},
	year={2020},
	publisher={APS}
}

@article{chen2020multi,
	title={Multi-layer Born multiple-scattering model for 3D phase microscopy},
	author={Chen, Michael and Ren, David and Liu, Hsiou-Yuan and Chowdhury, Shwetadwip and Waller, Laura},
	journal={Optica},
	volume={7},
	number={5},
	pages={394--403},
	year={2020},
	publisher={Optical Society of America}
}

@Article{schoenholz2020jax,
  author  = {Schoenholz, Samuel and Cubuk, Ekin Dogus},
  journal = {Adv. Neural Inf. Process. Syst.},
  title   = {Jax md: a framework for differentiable physics},
  year    = {2020},
  pages   = {11428--11441},
  volume  = {33},
}

@article{ali2021local,
	title={Local sound speed estimation for pulse-echo ultrasound in layered media},
	author={Ali, Rehman and Telichko, Arsenii V and Wang, Huaijun and Sukumar, Uday K and Vilches-Moure, Jose G and Paulmurugan, Ramasamy and Dahl, Jeremy J},
	journal={IEEE Trans. Ultrason. Ferroelectr. Freq. Control},
	volume={69},
	number={2},
	pages={500--511},
	year={2021},
	publisher={IEEE}
}

@article{lambert2022ultrasound2,
	title={Ultrasound matrix imaging—Part II: The distortion matrix for aberration correction over multiple isoplanatic patches},
	author={Lambert, William and Cobus, Laura A and Robin, Justine and Fink, Mathias and Aubry, Alexandre},
	journal={IEEE Trans. Med. Imaging},
	volume={41},
	number={12},
	pages={3921--3938},
	year={2022},
}

@Article{kamilov2023plug,
  author  = {Kamilov, Ulugbek S and Bouman, Charles A and Buzzard, Gregery T and Wohlberg, Brendt},
  journal = {IEEE Signal Process. Mag.},
  title   = {Plug-and-play methods for integrating physical and learned models in computational imaging: Theory, algorithms, and applications},
  year    = {2023},
  number  = {1},
  pages   = {85--97},
  volume  = {40},
}

@article{ali2023sound,
	title={Sound speed estimation for distributed aberration correction in laterally varying media},
	author={{Ali}, Rehman and {Mitcham}, Trevor M and {Singh}, Melanie and {Doyley}, Marvin M and {Bouchard}, Richard R and {Dahl}, Jeremy J and {Duric}, Nebojsa},
	journal={IEEE Trans. Comput. Imaging},
	volume={9},
	pages={367--382},
	year={2023},
}

@article{ali2023aberration,
	title={Aberration correction in diagnostic ultrasound: A review of the prior field and current directions},
	author={Ali, Rehman and Brevett, Thurston and Zhuang, Louise and Bendjador, Hanna and Podkowa, Anthony S and Hsieh, Scott S and Simson, Walter and Sanabria, Sergio J and Herickhoff, Carl D and Dahl, Jeremy J},
	journal={Z. fur Med. Phys.},
	volume={33},
	number={3},
	pages={267--291},
	year={2023},
	publisher={Elsevier}
}

@Article{bureau2023three,
  author    = {Bureau, Flavien and Robin, Justine and Le Ber, Arthur and Lambert, William and Fink, Mathias and Aubry, Alexandre},
  journal   = {Nat. Commun.},
  title     = {Three-dimensional ultrasound matrix imaging},
  year      = {2023},
  number    = {1},
  pages     = {6793},
  volume    = {14},
  publisher = {Nature Publishing Group UK London},
}

@Misc{bureau2023three_dataset,
	author       = {Bureau, Flavien and Robin, Justine and Le Ber, Arthur and Lambert, William and Fink, Mathias and Aubry, Alexandre},
	howpublished = {Zenodo},
	title        = {Three-Dimensional Ultrasound Matrix Imaging [Data set]},
	year         = {2023},
	doi          = {10.5281/zenodo.8159177},
}

@article{heriard2023refraction,
	title={Refraction-based speed of sound estimation in layered media: An angular approach},
	author={H{\'e}riard-Dubreuil, Baptiste and Besson, Adrien and Wintzenrieth, Fr{\'e}d{\'e}ric and Cohen-Bacrie, Claude and Thiran, Jean-Philippe},
	journal={IEEE Trans. Ultrason. Ferroelectr. Freq. Control},
	volume={70},
	number={6},
	pages={486--497},
	year={2023},
}

@article{stahli2023first,
	title={First-in-human diagnostic study of hepatic steatosis with computed ultrasound tomography in echo mode},
	author={St{\"a}hli, Patrick and Becchetti, Chiara and Korta Martiartu, Naiara and Berzigotti, Annalisa and Frenz, Martin and Jaeger, Michael},
	journal={Commun. Med.},
	volume={3},
	number={1},
	pages={176},
	year={2023},
}

@inproceedings{simson2023differentiable,
	title={Differentiable beamforming for ultrasound autofocusing},
	author={Simson, Walter and Zhuang, Louise and Sanabria, Sergio J and Antil, Neha and Dahl, Jeremy J and Hyun, Dongwoon},
	booktitle={International Conference on Medical Image Computing and Computer-Assisted Intervention},
	pages={428--437},
	year={2023},
	organization={Springer}
}

@Article{spainhour2024optimization,
  author  = {Spainhour, Jacob and Smart, Korben and Becker, Stephen and Bottenus, Nick},
  journal = {Phys. Med. Biol.},
  title   = {Optimization of array encoding for ultrasound imaging},
  year    = {2024},
  number  = {12},
  pages   = {125024},
  volume  = {69},
}

@Article{giraudat2024matrix,
  author  = {Giraudat, Elsa and Burtin, Arnaud and Le Ber, Arthur and Fink, Mathias and Komorowski, Jean-Christophe and Aubry, Alexandre},
  journal = {Commun. Earth Environ.},
  title   = {Matrix imaging as a tool for high-resolution monitoring of deep volcanic plumbing systems with seismic noise},
  year    = {2024},
  number  = {1},
  pages   = {509},
  volume  = {5},
}

@Article{balondrade2024multi,
  author  = {Balondrade, Paul and Barolle, Victor and Guigui, Nicolas and Auriant, Emeric and Rougier, Nathan and Boccara, Claude and Fink, Mathias and Aubry, Alexandre},
  journal = {Nat. Photonics},
  title   = {Multi-spectral reflection matrix for ultrafast 3D label-free microscopy},
  year    = {2024},
  number  = {10},
  pages   = {1097--1104},
  volume  = {18},
}

@article{garnier2025probing,
  title={Probing the speckle to estimate the effective speed of sound, a first step towards quantitative ultrasound imaging},
  author={Garnier, Josselin and Giovangigli, Laure and Goepfert, Quentin and Millien, Pierre},
  journal={arXiv preprint arXiv:2505.07566},
  year={2025}
}

@Article{simson2025ultrasound,
  author    = {Simson, Walter and Zhuang, Louise and Frey, Benjamin N and Sanabria, Sergio J and Dahl, Jeremy J and Hyun, Dongwoon},
  journal   = {IEEE Trans. Med. Imaging},
  title     = {Ultrasound autofocusing: Common midpoint phase error optimization via differentiable beamforming},
  year      = {2025},
  publisher = {IEEE},
}

@Article{schweizer2025pulse,
  author    = {Schweizer, Dieter and Farkas, Monika and Bezek, Can Deniz and Potempa, Anna and Leo, Cornelia and Kubik-Huch, Rahel A and Goksel, Orcun},
  journal   = {Ultrasound Med. Biol.},
  title     = {Pulse-echo imaging of breast speed of sound as a potential biomarker for breast cancer},
  year      = {2025},
  number    = {10},
  pages     = {1831-1839},
  volume    = {51},
  publisher = {Elsevier},
}

@Article{heriard2025path,
  author    = {H{\'e}riard-Dubreuil, Baptiste and Besson, Adrien and Cohen-Bacrie, Claude and Thiran, Jean-Philippe},
  journal   = {IEEE Trans. Med. Imaging},
  title     = {A Path-Based Model for Aberration Correction in Ultrasound Imaging},
  year      = {2025},
  number    = {8},
  pages     = {3222-3232},
  volume    = {44},
  publisher = {IEEE},
}

@Article{kang_tracing_2023,
  author    = {Kang, Sungsam and Kwon, Yongwoo and Lee, Hojun and Kim, Seho and Hong, Jin Hee and Yoon, Seokchan and Choi, Wonshik},
  journal   = {Nat. Commun.},
  title     = {Tracing multiple scattering trajectories for deep optical imaging in scattering media},
  year      = {2023},
  issn      = {2041-1723},
  month     = {oct},
  pages     = {6871},
  volume    = {14},
}

@Article{stoffa_splitstep_1990,
  author   = {Stoffa, P. L. and Fokkema, J. T. and De Luna Freire, R. M. and Kessinger, W. P.},
  journal  = {Geophysics},
  title    = {Split‐step Fourier migration},
  year     = {1990},
  issn     = {0016-8033, 1942-2156},
  month    = apr,
  number   = {4},
  pages    = {410--421},
  volume   = {55},
}

@Article{montaldo_coherent_2009,
  author   = {Montaldo, Gabriel and Tanter, Mickaël and Bercoff, Jérémy and Benech, Nicolas and Fink, Mathias},
  journal  = {IEEE Trans. Ultras. Ferroel. Freq. Cont.},
  title    = {Coherent plane-wave compounding for very high frame rate ultrasonography and transient elastography},
  year     = {2009},
  issn     = {1525-8955},
  month    = mar,
  number   = {3},
  pages    = {489--506},
  volume   = {56},
  pmid     = {19411209},
}

@article{lambert_distortion_2020,
	title = {Distortion matrix approach for ultrasound imaging of random scattering media},
	volume = {117},
	issn = {0027-8424, 1091-6490},
	number = {26},
	urldate = {2021-12-23},
	journal = {Proceedings of the National Academy of Sciences},
	author = {Lambert, William and Cobus, Laura A. and Frappart, Thomas and Fink, Mathias and Aubry, Alexandre},
	month = {jun},
	year = {2020},
	pages = {14645--14656},
}

@Article{Devaney1982,
  author       = {Devaney, A},
  journal      = {Ultrason. Imaging},
  title        = {A filtered backpropagation algorithm for diffraction tomography},
  year         = {1982},
  pages        = {336--350},
  volume       = {4},
  date         = {1982},
  journaltitle = {Ultrason. Imaging},
  publisher    = {Elsevier BV},
}

@InBook{Thurber2015,
  author    = {Thurber, C. and Ritsema, J.},
  pages     = {307--337},
  publisher = {Elsevier},
  title     = {Theory and Observations - Seismic Tomography and Inverse Methods},
  year      = {2015},
  booktitle = {Treatise on Geophysics},
}

@Article{Simson2024,
  author       = {Simson, W. A. and Paschali, M. and Sideri-Lampretsa, V. and Navab, N. and Dahl, J. J.},
  journal      = {Ultrasonics},
  title        = {Investigating pulse-echo sound speed estimation in breast ultrasound with deep learning},
  year         = {2024},
  pages        = {107179},
  volume       = {137},
  date         = {2024},
  journaltitle = {Ultrasonics},
}

@Article{Wolf1969,
  author    = {Wolf, E.},
  journal   = {Opt. Commun.},
  title     = {Three-dimensional structure determination of semi-transparent objects from holographic data},
  year      = {1969},
  month     = sep,
  pages     = {153--156},
  volume    = {1},
  publisher = {Elsevier BV},
}

@Article{badon_distortion_2020,
  author       = {Badon, Amaury and Barolle, Victor and Irsch, Kristina and Boccara, A. Claude and Fink, Mathias and Aubry, Alexandre},
  journal      = {Sci. Adv.},
  title        = {Distortion matrix concept for deep optical imaging in scattering media},
  year         = {2020},
  issn         = {2375-2548},
  month        = jul,
  number       = {30},
  pages        = {aay7170},
  volume       = {6},
  date         = {2024},
  journaltitle = {Nature Photon. (in press)},
  publisher    = {American Association for the Advancement of Science (AAAS)},
}

@Article{najar_non-invasive_2023,
  author       = {Najar, U. and Barolle, V. and Balondrade, P. and Fink, M. and Boccara, A. C. and Fink, M. and Aubry, A.},
  journal      = {Nat. Commun.},
  title        = {Harnessing Forward Multiple Scattering for Optical Imaging Deep Inside an Opaque Medium},
  year         = {2024},
  pages        = {7349},
  volume       = {15},
  howpublished = {arXiv:2303.06119},
}

@Article{Yoon2020,
  author    = {Yoon, Seokchan and Kim, Moonseok and Jang, Mooseok and Choi, Youngwoon and Choi, Wonjun and Kang, Sungsam and Choi, Wonshik},
  journal   = {Nat. Rev. Phys.},
  title     = {Deep optical imaging within complex scattering media},
  year      = {2020},
  issn      = {2522-5820},
  month     = feb,
  number    = {3},
  pages     = {141--158},
  volume    = {2},
  publisher = {Springer Science and Business Media LLC},
}

@Article{Villiger2016,
  author    = {Villiger, Martin and Lorenser, Dirk and McLaughlin, Robert A. and Quirk, Bryden C. and Kirk, Rodney W. and Bouma, Brett E. and Sampson, David D.},
  journal   = {Sci. Rep.},
  title     = {Deep tissue volume imaging of birefringence through fibre-optic needle probes for the delineation of breast tumour},
  year      = {2016},
  issn      = {2045-2322},
  month     = jul,
  number    = {1},
  pages     = {28771},
  volume    = {6},
  doi       = {10.1038/srep28771},
  publisher = {Springer Science and Business Media LLC},
}

@Article{Tam1981,
  author    = {Tam, K. C. and Perez-Mendez, V.},
  journal   = {J. Opt. Soc. Am.},
  title     = {Tomographical imaging with limited-angle input},
  year      = {1981},
  issn      = {0030-3941},
  month     = may,
  number    = {5},
  pages     = {582},
  volume    = {71},
  publisher = {Optica Publishing Group},
}

@Article{Osnabrugge2016,
  author    = {Osnabrugge, Gerwin and Leedumrongwatthanakun, Saroch and Vellekoop, Ivo M.},
  journal   = {J. Comput. Phys.},
  title     = {A convergent Born series for solving the inhomogeneous Helmholtz equation in arbitrarily large media},
  year      = {2016},
  issn      = {0021-9991},
  month     = oct,
  pages     = {113--124},
  volume    = {322},
  publisher = {Elsevier BV},
}

@Article{Haim2024,
  author    = {Haim, Omri and Boger-Lombard, Jeremy and Katz, Ori},
  journal   = {Nat. Photonics},
  title     = {Image-guided computational holographic wavefront shaping},
  year      = {2024},
  issn      = {1749-4893},
  month     = oct,
  number    = {1},
  pages     = {44--53},
  volume    = {19},
  publisher = {Springer Science and Business Media LLC},
}

@InBook{Duck1990,
  author    = {Duck, Francis A.},
  pages     = {73--135},
  publisher = {Elsevier},
  title     = {Acoustic Properties of Tissue at Ultrasonic Frequencies},
  year      = {1990},
  isbn      = {9780122228001},
  booktitle = {Physical Properties of Tissues},
}

@Article{Li2025,
  author    = {Li, Tongyu and Zhu, Jiabei and Shen, Yi and Tian, Lei},
  journal   = {Optica},
  title     = {Reflection-mode diffraction tomography of multiple-scattering samples on a reflective substrate from intensity images},
  year      = {2025},
  issn      = {2334-2536},
  month     = mar,
  number    = {3},
  pages     = {406},
  volume    = {12},
  publisher = {Optica Publishing Group},
}

@Article{Lan2014,
  author    = {Lan, B. and Lowe, M. and Dunne, F.P.E.},
  title     = {Experimental and computational studies of ultrasound wave propagation in hexagonal close-packed polycrystals for texture detection},
  doi       = {10.1016/j.actamat.2013.10.012},
  issn      = {1359-6454},
  pages     = {107--122},
  volume    = {63},
  journal   = {Acta Materialia},
  month     = jan,
  publisher = {Elsevier BV},
  year      = {2014},
}

@Article{Ranganathan2008,
  author    = {Ranganathan, Shivakumar I. and Ostoja-Starzewski, Martin},
  journal   = {Phys. Rev. Lett.},
  title     = {Universal Elastic Anisotropy Index},
  year      = {2008},
  number    = {5},
  pages     = {055504},
  volume    = {101},
  publisher = {American Physical Society (APS)},
}

@Article{Bertolotti2022,
  author    = {Bertolotti, Jacopo and Katz, Ori},
  journal   = {Nature Physics},
  title     = {Imaging in complex media},
  year      = {2022},
  issn      = {1745-2481},
  month     = sep,
  number    = {9},
  pages     = {1008--1017},
  volume    = {18},
  doi       = {10.1038/s41567-022-01723-8},
  publisher = {Springer Science and Business Media LLC},
}

@Article{Judkewitz2015,
  author    = {Judkewitz, Benjamin and Horstmeyer, Roarke and Vellekoop, Ivo M. and Papadopoulos, Ioannis N. and Yang, Changhuei},
  journal   = {Nature Physics},
  title     = {Translation correlations in anisotropically scattering media},
  year      = {2015},
  issn      = {1745-2481},
  month     = jun,
  number    = {8},
  pages     = {684--689},
  volume    = {11},
  doi       = {10.1038/nphys3373},
  publisher = {Springer Science and Business Media LLC},
}

@article{long2018lag,
	title={Lag-one coherence as a metric for ultrasonic image quality},
	author={Long, Will and Bottenus, Nick and Trahey, Gregg E},
	journal={IEEE transactions on ultrasonics, ferroelectrics, and frequency control},
	volume={65},
	number={10},
	pages={1768--1780},
	year={2018},
	publisher={IEEE}
}

@Article{Estrada2018,
  author    = {Estrada, Héctor and Huang, Xiao and Rebling, Johannes and Zwack, Michael and Gottschalk, Sven and Razansky, Daniel},
  title     = {Virtual craniotomy for high-resolution optoacoustic brain microscopy},
  doi       = {10.1038/s41598-017-18857-y},
  issn      = {2045-2322},
  number    = {1},
  pages     = {1459},
  volume    = {8},
  journal   = {Sci. Rep.},
  publisher = {Springer Science and Business Media LLC},
  year      = {2018},
}

@Article{White2006,
  author    = {White, P.J. and Clement, G.T. and Hynynen, K.},
  title     = {Longitudinal and shear mode ultrasound propagation in human skull bone},
  doi       = {10.1016/j.ultrasmedbio.2006.03.015},
  issn      = {0301-5629},
  number    = {7},
  pages     = {1085--1096},
  volume    = {32},
  journal   = {Ultras. Med. Biol.},
  publisher = {Elsevier BV},
  year      = {2006},
}

@Article{Bureau2024,
  author   = {Bureau, Flavien and Brenner, Emma and Korta Martiartu, Naiara Giraudat, Elsa and Le Ber, Arthur and Lambert, William and Carmier, Louis and Guibal, Aymeric and Fink, Mathias and Aubry, Alexandre},
  journal  = {Phys. Rev. Appl. (accepted)},
  title    = {Self-Portrait of the Focusing Process in Speckle: {II.} {G}ouy Phase Shift for Defocus Correction and Pixel Depth Reassignment},
  year     = {2026},
}

@Misc{heriard2026,
  author       = {H{\'e}riard-Dubreuil, Baptiste and Brenner, Emma and Rio, Benjamin and Lambert, William and Chamming's, Foucauld and Fink, Mathias and Aubry, Alexandre},
  howpublished = {Zenodo},
  note         = {https://doi.org/10.5281/zenodo.23419267},
  title        = {Physics-Based Learning of the Wave Speed Landscape in Complex Media [data]},
  year         = {2026},
}

\clearpage 

\noindent\textbf{Acknowledgments.}
We wish to thank F. Bureau for fruitful discussions on the method and its applications as well as for providing the ultrasound data of the 3D experiment. We also wish to thank the authors of Ref.\cite{ali2023sound} for publishing in open-access the experimental ultrasound data used for the 2D in vitro experiment.

\vspace{2 mm}
\noindent\textbf{Funding.}
The authors are grateful for the funding provided by Labex WIFI (Laboratory of Excellence within the French Program Investments for the Future; ANR-10-LABX-24 and ANR-10-IDEX-0001-02 PSL*). This project has also received funding from the European Research Council (ERC) under the European Union’s Horizon 2020 research and innovation program (grant agreements no. 819261 and 101292719, REMINISCENCE and DIONISOS projects, respectively, A.A.). E.B. acknowledges financial support from the SuperSonic Imagine company. 

\vspace{2 mm}

\noindent\textbf{Author Contributions Statement.}
B.H.-D. and A.A. initiated the project. B.H.-D. and A.A. conceived the method. B.H.-D and A.A. developed the theoretical background. E.B. developed the implementation of the split step Fourier method. B.H.-D. implemented the differential matrix imaging process.  F.C. conducted the breast tissue experiment. {E.B. and W.L. gathered and processed the breast tissue data}. B.R. adapted the method to this configuration.  B.H.-D. and A.A. prepared the manuscript. B.H.-D., E.B., B.R., W.L., F.C., M.F. and A.A. contributed to the discussion of the results and the development of the manuscript. 

\vspace{2 mm}

\noindent\textbf{Competing interests.}
B.H.-D., M.F. and A.A. are inventors on a patent related to this work (no. FR2510672, filled in September 2025). E.B. and W.L. were employees of the SuperSonic Imagine company. All authors declare that they have no other competing interests.

\clearpage

\renewcommand{\thetable}{\arabic{table}}
\renewcommand{\thefigure}{\arabic{figure}}
\renewcommand{\theequation}{S\arabic{equation}}
\renewcommand{\thesection}{\arabic{section}}
\renewcommand{\thesubsection}{\arabic{section}.\arabic{subsection}}
\renewcommand{\figurename}{Supplementary Fig.}

\setcounter{equation}{0}
\setcounter{table}{0}

\setcounter{page}{1}

\newpage

\title{Physics-Based Learning of the Wave Speed Landscape in Complex Media}

\author{Baptiste~H\'eriard-Dubreuil}
\thanks{Corresponding authors. Email: baptiste.heriard-dubreuil@espci.fr, alexandre.aubry@espci.fr}
\affiliation{Institut Langevin, ESPCI, PSL University, CNRS, Paris, France.}
\author{Emma~Brenner}
\affiliation{Institut Langevin, ESPCI, PSL University, CNRS, Paris, France.}
\affiliation{SuperSonic Imagine Company, Aix-en-Provence, France.}
\author{Benjamin~Rio}
\affiliation{Institut Langevin, ESPCI, PSL University, CNRS, Paris, France.}
\author{William~Lambert}
\affiliation{SuperSonic Imagine Company, Aix-en-Provence, France.}
\author{Foucauld~Chamming's}
\affiliation{Institut Bergoni\'e, Bordeaux, France.}
\author{Mathias~Fink}
\affiliation{Institut Langevin, ESPCI, PSL University, CNRS, Paris, France.}
\author{Alexandre Aubry}
\thanks{Corresponding authors. Email: baptiste.heriard-dubreuil@espci.fr, alexandre.aubry@espci.fr}
\affiliation{Institut Langevin, ESPCI, PSL University, CNRS, Paris, France.}

\maketitle

\counterwithin*{figure}{part}
\counterwithin*{table}{part}

\stepcounter{part}

\section*{Table of Contents}

\begin{itemize}
\item \ref{sup:fact}. Factorization of the Reflection Matrix
\item \ref{sup:diff}. Differentiability and Gradient implementation
\item \ref{sup:tv_filt}. Local Minima, Regularization and Filtering
\item \ref{sup:relax}. Constraint Relaxation
\item \ref{sec:simulation}. Simulation Validation
\item \ref{sec:parameters}. Parameter Study
\item \ref{sec:acquisition_effect}. Effect of Acquisition
\item \ref{sec:sota}. Comparison with State of the Art Methods
\item \ref{sec:complexity}. Computation Time and Memory Requirements
\item \ref{sec:3d_extended}. Extended results
\item Captions of Supplementary Videos 1-7
\end{itemize}

\clearpage

\renewcommand{\thesection}{Supplementary Note \arabic{section}}

\section{Factorization of the Reflection Matrix}
\label{sup:fact}

The most expensive operation of the proposed algorithm is the iteration of equation~\eqref{eq:split_step_iteration}. It can be implemented without any matrix multiplication using only separable fast Fourier transforms and element-wise multiplications, leading to an overall computation cost at each depth for each frequency in the order of $\mathcal{O}\left(N_x^2\log N_x\right)$ in 2D and $\mathcal{O}\left(N_x^2 N_y^2\log N_x \right)$ in 3D (taking $N_x > N_y$ without loss of generality).

In ultrasound imaging, the number of emissions is usually significantly lower than the number of lateral positions $N_i \ll N_{x}$. To leverage this property, we replace focused reflection matrices of size $N_x \times N_x$ by two matrices of size $N_i \times N_x$ (for each depth and each frequency). These two matrices, denoted as $\mathbf{T}_{\mathbf{i}\mathbf{x}}(z, f; c)$ and $\mathbf{R}_{\mathbf{i}\mathbf{x}}(z, f; c)$, correspond to the emission part and the reception part of the focused reflection matrices. They can be computed recursively in a similar way as the focused reflection matrices. Initialization is performed using the interpolation matrices as in equation~\eqref{eq:split_step_init}:
\begin{equation} 
    \begin{cases}
    \mathbf{T}_{\mathbf{i}\mathbf{x}}(0, f; c) = \mathbf{I}^\dagger_{\mathbf{x}\mathbf{i}}(f; c), \\
   \mathbf{R}_{\mathbf{i}\mathbf{x}}(0, f; c) = \mathbf{R}_{\mathbf{i}\mathbf{o}}(f) \times \mathbf{I}^T_{\mathbf{x}\mathbf{o}}(f; c).
    \end{cases}
    \label{eq:fact_init}
\end{equation}

\noindent
Their propagation from depth $z$ to $z + \delta z$ is performed with the following update formulae:
\begin{equation}
    \begin{cases}
    \mathbf{T}_{\mathbf{i}\mathbf{x}}(z + \delta z, f; c) = \mathbf{T}_{\mathbf{i}\mathbf{x}}(z + \delta z, f; c)\times  \mathbf{\bm{\delta} T}^T(\delta z, f; c(z)), \\
    \mathbf{R}_{\mathbf{i}\mathbf{x}}(z + \delta z, f; c) = \mathbf{R}_{\mathbf{i}\mathbf{x}}(z + \delta z, f; c)\times  \mathbf{\bm{\delta} T}^\dagger(\delta z, f; c(z)).
    \end{cases}
    \label{eq:fact_update}
\end{equation}

\noindent
With this representation, we can recover the focused reflection matrices at each depth:
\begin{equation}
   \mathbf{R}_{\mathbf{x}\mathbf{x}}(z, f; c) =\mathbf{T}^{\dagger}_{\mathbf{i}\mathbf{x}}(z, f; c) \times  \mathbf{R}_{\mathbf{i}\mathbf{x}}(z, f; c).
    \label{eq:fact2r}
\end{equation}
These two matrices therefore contain all the information of the focused reflection matrices but are computationally significantly cheaper due to their reduced size. They result in a computation cost at each depth for each frequency of $\mathcal{O}\left(N_i N_x\log N_x\right)$ in 2D and $\mathcal{O}\left(N_i N_x N_y\log N_x \right)$ in 3D. 
However, the matrix multiplication of equation~\eqref{eq:fact2r} is expensive and becomes the bottleneck of our algorithm. To circumvent this issue, we rather work with the distortion matrix~\cite{lambert2022ultrasound2}, obtained by element-wise multiplication of our two matrices:
\begin{equation}
    \mathbf{D}_{\mathbf{i}\mathbf{x}}(z, f; c) =\mathbf{T}^{*}_{\mathbf{i}\mathbf{x}}(z, f; c) \odot  \mathbf{R}_{\mathbf{i}\mathbf{x}}(z, f; c).
\end{equation}
One can notice that the diagonal of the focused reflection matrix, equal to the confocal reflectivity image, can be recovered directly from the distortion matrix by summing its rows:
\begin{equation}
    \mathbf{I}(z; c) = \sum\limits_f \sum\limits_{\mathbf{i}}
    \mathbf{D}_{\mathbf{i}\mathbf{x}}(z, f; c).
\end{equation}
As a result, we can compute the numerator of the focusing quality index without explicitly computing the focused reflection matrices. However, the denominator cannot be obtained without a large amount of multiplications. We use in practice an approximation of this denominator by computing the norm of the distortion matrix. The obtained focusing quality index $\widetilde{\mathcal{F}}(c)$ resembles the focusing criterion, the so-called coherence factor, introduced by Mallart and Fink~\cite{mallart1994adaptive}:
\begin{equation}
    \widetilde{\mathcal{F}}(c) = \left\langle \frac{|\sum\limits_{f, \mathbf{i}}\mathcal{D}(\mathbf{i}, \mathbf{x}, z, f; c)|^2}{ \sum\limits_{\mathbf{i}}|\sum\limits_{f}\mathcal{D}(\mathbf{i}, \mathbf{x}, f, z; c)|^2} \right\rangle_{\mathbf{x}, z}.
    \label{eq:focusing_quality_d}
\end{equation}
With this version of the quality factor, the focused reflection matrices is never explicitly computed and the computational complexity and memory footprint is therefore kept low.

This factorization results in a large gain in computation time and memory requirement, providing that the number of emissions is significantly lower than the number of lateral positions. In particular, this simplification is key to generalize to 3D configurations.

\section{Differentiability and Gradient implementation}
\label{sup:diff}

We want to compute the gradient of the focusing quality factor with respect to the velocity model to perform an optimization, as represented by the lower row of Figure~\ref{fig:method} in the accompanying paper. This gradient can be computed using the chain rule as in equation~\eqref{eq:gradient}. In more details, we obtain:
\begin{equation}
    \frac{\partial \mathcal{F}}{\partial c} = \sum\limits_{z, f}\frac{\partial \mathcal{F}}{\partial \mathbf{R}(z, f; c)} \sum\limits_{z' \leq z} \frac{\partial \mathbf{R}(z, f; c)}{\partial \mathbf{\bm{\delta} T}(\delta z, f; c(z'))} \frac{\partial \mathbf{\bm{\delta} T}(\delta z, f; c(z'))}{\partial c},
    \label{eq:gradient_detailed}
\end{equation}
where the subscript $-_{\mathbf{x}\mathbf{x}}$ has been removed for the sake of clarity.

In our case, the chain rule is performed automatically by Pytorch.
Nevertheless, the study of the differentiability of all operators and of their derivatives is crucial for a robust and efficient optimization.
In our case, the most complex operation is the computation of $\mathbf{\bm{\delta} T}$ which depends on the wave velocity $c$ via the two vectors $\mathbf{h}$ and $\mathbf{s}$, as detailed in equations~\eqref{eq:free_space_prop} and~\eqref{eq:phase_screen}.
In these two equations, vectors $\mathbf{h}$ and $\mathbf{s}$ depend on the slowness $\sigma=1/c$ rather than the wave speed $c$. We therefore choose to optimize the focusing quality with respect to the slowness rather than the wave velocity.
In addition, the expression $\mathbf{h}$ is the only non-differentiable operation as it involves a complex square root which can be null for some frequencies. To remove the singularity at zero and bound the derivative of our model, we modify the formula of $\mathbf{h}$ to include a small complex offset, resulting in a smooth expression in the complex plane:
\begin{equation}
    \widetilde{\mathbf{h}}(f, c(z)) = \exp{\left(j 2\pi \delta z \sqrt{\left(f \sigma_m(z) \right)^2 - |\mathbf{f_{x}}|^2 +
     j\epsilon}\right)},
\end{equation}
with $\epsilon$ a small positive real number, usually set to $\epsilon=|df_x|^2$ with $df_x$ the spatial frequency step.

Finally, the gradient computation requires a large amount of memory, mainly due to the first two terms of equation~\eqref{eq:gradient_detailed}. Indeed, the focusing quality depends on all the focused reflection matrices (i.e. for all depths and frequencies), which themselves depend on all previous elementary transmission operators $\mathbf{\bm{\delta} T}(\delta z, f; c(z'))$ for $z'<z$.
Even with a clever implementation, this gradient computation requires to store all reflection matrices (or equivalently all elementary transmission matrices) as intermediate values. To limit the memory footprint of the gradient, we separate its computation into a first part where frequencies are handled independently, and a second part that mixes the frequencies. We compute the gradient of the first part by batching on the frequencies, and combine all results with the second part. This strategy gives a new parameter (the batch size) which controls the trade-off between computation efficiency and memory management.

\section{Local Minima, Regularization and Filtering}
\label{sup:tv_filt}

When we apply the naive gradient ascent algorithm on a simulation of a heterogeneous medium, we obtain a result that shows high short-scale spatial variations, as represented in~Supplementary Figure~\ref{fig:sup_reg}a.
These variations do not correspond to any physical reality and are interpreted as overfitting.
Since we seek to recover the large-scale variations of the wave velocity, we introduce a smoothness prior in the form of a regularization to mitigate such variations:
\begin{equation}
	\max\limits_{\sigma} \mathcal{F}(\sigma) - \eta TV(\sigma),
	\label{eq:opt_pb_reg}
\end{equation}
with $TV(\sigma) = \left\langle || \nabla \sigma||_1 \right\rangle_{x, z}$ a total variation regularization implemented in its anisotropic \mbox{$\ell$-1} version and $\eta$ a regularization weight.
The result is displayed in Supplementary Figure~\ref{fig:sup_reg}b. and shows a drastic reduction of noise in the velocity map.
\begin{figure*} 
	\centering
	\includegraphics[width=\textwidth]{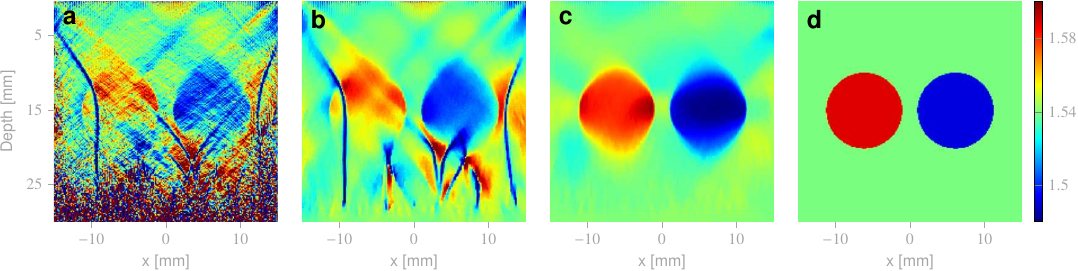} 

	\caption{\textbf{Effect of regularization and filtering on the convergence.}
	Wave velocity maps at convergence for an ultrasound simulation. \textbf{a}, Result without regularization nor filtering. \textbf{b}, Result with regularization but without filtering. \textbf{c}, Result with regularization and filtering. \textbf{d}, Ground truth. All speed of sound values are given in mm.\textmu s$^{-1}$.}
	\label{fig:sup_reg}
\end{figure*}
However, a few short-scale spatial variations remain. Such artifacts are particularly stable as they are often associated with delays of exactly one or several cycles of the emitted pulse. They are described as cycle skipping artifacts in the tomography literature and correspond to phase interference of waves.
To avoid this pitfall without enforcing an unreasonable smoothness prior on the slowness map, we perform a median filtering step at each iteration on the result of the gradient ascent. The gradient ascent steps becomes $\sigma\leftarrow f(\sigma + \epsilon (\nabla_\sigma \mathcal{F}(\sigma) - \eta \nabla_\sigma TV(\sigma)))$, where $f$ is a partial median filer defined as:
\begin{equation}
	f(\sigma) = \alpha \cdot \sigma + (1-\alpha)\cdot m(\sigma),
\end{equation}
with $m(-)$ a median filtering operator and $\alpha$ a small constant usually set to $\alpha=0.5$.
This filtering step can be interpreted as a denoising operation similar to the implicit proximal of plug and play algorithms \cite{kamilov2023plug}.
It results in an accurate speed-of-sound estimator (see Supplementary Figure~\ref{fig:sup_reg}) and accelerates the overall convergence.

\section{Constraint Relaxation}
\label{sup:relax}
If the regularization and the filtering solve most convergence issues, an asymmetry of the forward problem remains.
Indeed, the focusing quality is affected by phase aberrations accumulated during wave propagation. This accumulation corresponds to axial integrals of the slowness map, resulting in difficulties to recover the high-frequency axial variations of the wave velocity. This limitation is linked to the so-called missing cone problem in tomography \cite{Tam1981}. In our case, the computation of the focusing quality at all depths and the presence of scattering and refraction effects mitigate the missing cone problem and enables us to accurately recover an axial profile.
However, this asymmetry still leads to a slow axial convergence compared to the estimation of lateral variations.

As a result, our algorithm tends to compensate axial velocity variations by creating an artificial lateral profile, as shown in Supplementary Figure~\ref{fig:sup_sep}b.
\begin{figure*}
		\centering
		\includegraphics[width=\textwidth]{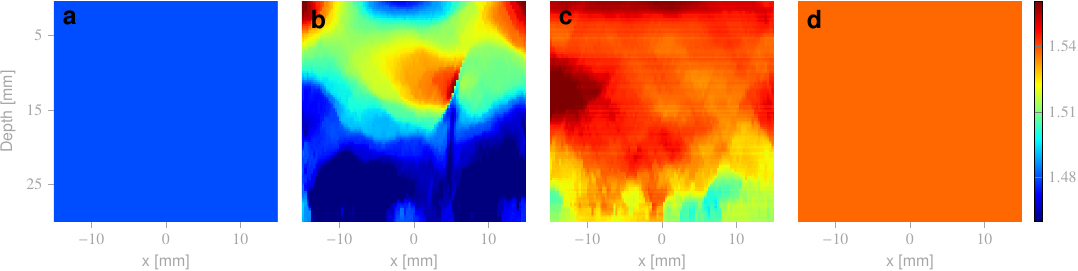} 
        
		\caption{\textbf{Effect of variable splitting on the convergence.}
			Wave velocity maps at convergence for an ultrasound simulation. \textbf{a}, Initial hypothesis for the wave velocity. \textbf{b}, Result without variable splitting. \textbf{c}, Result with variable splitting. \textbf{d}, Ground truth. All speed of sound values are given in mm.\textmu s$^{-1}$.}
		\label{fig:sup_sep}
	\end{figure*}
The mathematical origin of this effect takes its roots in the split-step Fourier algorithm in which the wave slowness is separated into an axial profile term $\sigma_m$ used for the free-space propagation in equation~\eqref{eq:free_space_prop} and a residual lateral term $\delta \sigma$ used as a phase mask in equation~\eqref{eq:phase_screen}.
This separation may result in contrary gradient directions which hinder the convergence.
To circumvent this issue, we separate the slowness into two different variables $\sigma_m$ and $\delta\sigma$, used for the free-space propagation (equation~\eqref{eq:free_space_prop}) and the phase screens (equation~\eqref{eq:phase_screen}) respectively. Our optimization problem becomes:
\begin{equation}
	\begin{aligned}
		\max\limits_{\delta \sigma, \sigma_m} \quad & \mathcal{F}(\delta \sigma, \sigma_m) - \eta TV(\delta \sigma, \sigma_m), \\
		\textrm{such that} \quad & \left\langle\delta \sigma\right\rangle_{\mathbf{x}} = 0.
	\end{aligned}
	\label{eq:opt_pb_constraint}
\end{equation}
In this form, the constraint may still block the gradient ascent. We perform a Lagrangian relaxation of this constraint to resolve the conflict:
\begin{equation}
		\max\limits_{\delta \sigma, \sigma_m} \mathcal{F}(\delta \sigma, \sigma_m) - \eta TV(\delta \sigma, \sigma_m) - \mu || \left\langle\delta \sigma\right\rangle_{x} ||,
	\label{eq:opt_pb_relaxed}
\end{equation}
where $||-||$ can be any norm. In our case, we take $\mu=0$, leading to a loss similar to the one of equation \eqref{eq:opt_pb_reg}, but optimized on two independent variables.
Such a relaxation can be interpreted as a dimension increase that facilitates the convergence.

Finally, we adapt the gradient ascent by considering two optimization phases. In the first phase, we only optimize the axial profile $\sigma_m$. In the second phase, we start from the result of the first phase and optimize jointly the two variables.
The variable splitting combined with this adapted ascent algorithm results in an accurate estimation of both the axial profile of the wave velocity and its lateral variations, as seen in Supplementary Figure~\ref{fig:sup_sep}c.

\section{Simulation Validation}
\label{sec:simulation}

We validate the proposed method using simulations performed with k-Wave~\cite{treeby2010k}. Simulation parameters are reported in Tab.~\ref{tab:parameters} and results are displayed in Fig.~\ref{fig:sup_simu} of the accompanying paper.

For each simulation, we define a wave velocity map made of large scale variations such as inclusions or layers, and a density map containing short scale variations such as sub-wavelength scatterers and anechoic regions.
We simulate ultrasound acquisitions in such media, and apply the proposed method to the resulting signals. The optimization is performed on one NVIDIA RTX 6000 Ada Generation GPU, converging in 1 to 35 minutes (depending on the grid size).

We display in Fig.~\ref{fig:sup_simu} the aberrated reflection images obtained with a uniform wave velocity hypothesis of $1540$~m.s$^{-1}$ and compare it with the reflection image obtained at convergence. In all cases, a sharp reflection image is recovered without any geometric distortions, which validates the ability to correct aberrations with the proposed method.

The estimated wave velocity map is compared to the simulated one on the right part of Fig.~\ref{fig:sup_simu}.
In the three first simulations, we obtain an accurate velocity estimate with a high lateral resolution (approximately one wavelength) and an axial resolution of a few wavelengths. This asymmetric resolution is a direct manifestation of the missing cone problem even though it is here partially mitigated by refraction and scattering. Obtained velocity values are close to the ground truth, with absolute errors lower than 20~m.s$^{-1}$.

The last simulation models a more complex configuration. In this case, we observe a good agreement between the estimated and ground truth value in the center of the domain. However, we observe some lateral border effects as well as a low accuracy at high depths. These limitations can be explained by the low impact of the velocity values at such positions on the average focusing quality.
For future applications, results could be improved by masking or correcting such border effects. 

\section{Parameter Study}\label{sec:parameters}

In this section, we study the effect of reconstruction parameters on the resulting wave velocity map. We perform this study on the simulation dataset described in Supplementary Section~\ref{sec:simulation}, consisting in a homogeneous background at 1540~m.s$^{-1}$ with two circular inclusions of $\pm$50~m.s$^{-1}$.

We study the effect of the initial velocity guess, the grid resolution, the learning rate and the regularization weight. Unless stated otherwise, we use the parameters described in the first column of Table~\ref{tab:parameters}, with a domain size restricted to 100$\times$100 wavelengths for computational considerations.

\subsection{Initial Velocity Guess}

To study the effect of the initial velocity guess, we modify the initialization of the optimization scheme. In order to be blind to the true velocity background, we define the reconstructed grid to be 100$\times$100 wavelengths, the wavelength $\lambda$ being obtained directly from the initial velocity guess as $\lambda = c_{\text{init}} / f_c$, with $f_c$ the central frequency. Reconstructed velocity maps obtained for various $c_{\text{init}}$, together with their mean absolute error with respect to the true wave velocity map are displayed in Supplementary Fig.~\ref{fig:sup_initial_sos}. They show that the initial wave velocity guess has almost no impact on the upper part of the reconstructed map, but creates a bias in the deepest 10~mm.

\begin{figure*}[htp]
	\includegraphics[width=\textwidth]{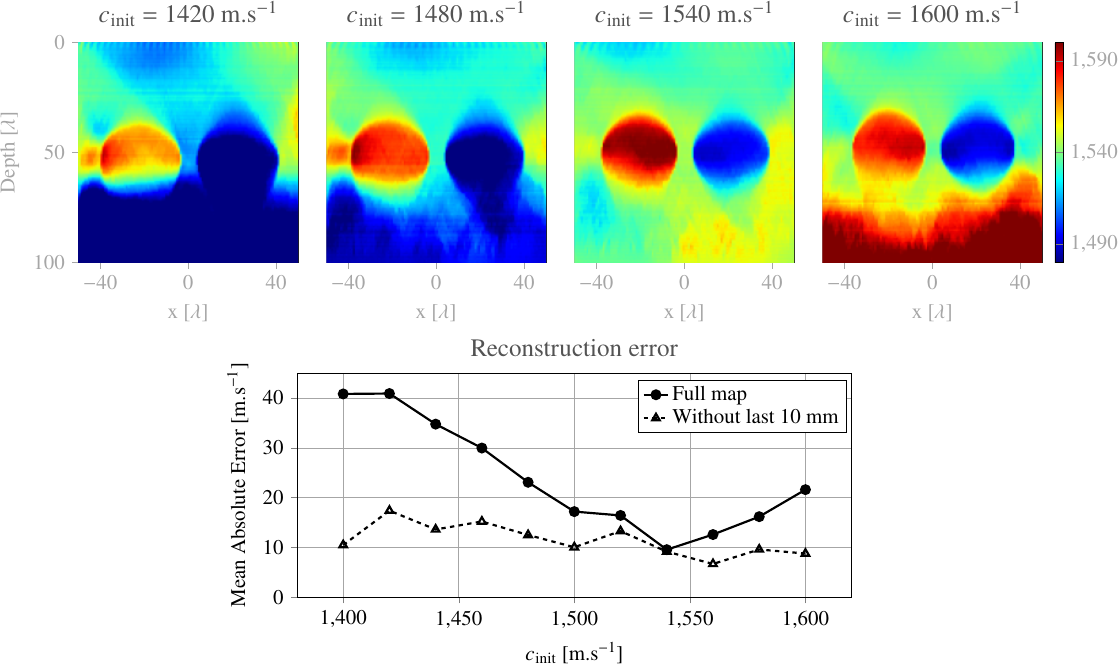}
	\caption{\textbf{Effect of initial velocity guess}. Top: reconstructed wave velocity map for different initial guesses. Bottom: Mean absolute error, compute on the full map (solid line) or removing the last 10~mm (dashed line).}\label{fig:sup_initial_sos}
\end{figure*}

This effect can be understood from a tomographic point of view. Indeed, the wave velocity at a given point affects the focusing quality only at points located below it. As a result, the wave velocity associated to the deepest points of the reconstructed domain have almost no effect on the overall focusing quality and are therefore not updated during the gradient ascent process. They remain at their initial value $c_{\text{init}}$.

Such a limitation can be avoided with various techniques. 
For instance, one could adopt a regularization weight varying with depth in order to integrate the tomographic perspective described in the previous paragraph directly in the optimization scheme.
Another solution, described for instance in \cite{simson2025ultrasound}, would consist in starting the optimization with a preliminary round dedicated to finding an optimal constant wave velocity before running the full optimization.
Finally, as the focusing quality weakly depends on the wave velocity at high depth, we could dissociate the domain used for the loss estimation and the one used for the wave velocity reconstruction, e.g. cropping the last 10~mm of the obtained velocity map.

\subsection{Reconstruction Resolution}

	The experiments performed in this work all use spatial grids with an isotropic step chosen equal to or slightly smaller than the wavelength ($\delta x = \delta z = $ 1 or 0.8$\lambda$).
	However, experiments in optics and ultrasound have shown that aberrations caused by wave velocity variations can be corrected with only a few phase screens \cite{ali2023aberration, kang_tracing_2023, Haim2024}.
	This observation raises the question of the optimal reconstruction grid for wave velocity imaging, where we want to not only correct aberrations but also recreate a faithful estimates of the wave velocity landscape.

	To study the effect of the reconstruction resolution on the obtained velocity map, we test the proposed method on a simulation and vary independently the axial and lateral resolution of the considered grid. Resulting wave velocity maps are displayed in Supplementary~Fig.~\ref{fig:sup_res}, for $\delta x=\lambda/2$ and a varying $\delta z \in [\lambda/2, 4\lambda]$; and for $\delta z=\lambda/2$ and a varying $\delta x \in [\lambda/2, 4\lambda]$.
	As per the previous section, we also compute the mean absolute error with respect to the true velocity map for all pairs of axial and lateral resolutions.

\begin{figure*}[htp]
	\includegraphics[width=\textwidth]{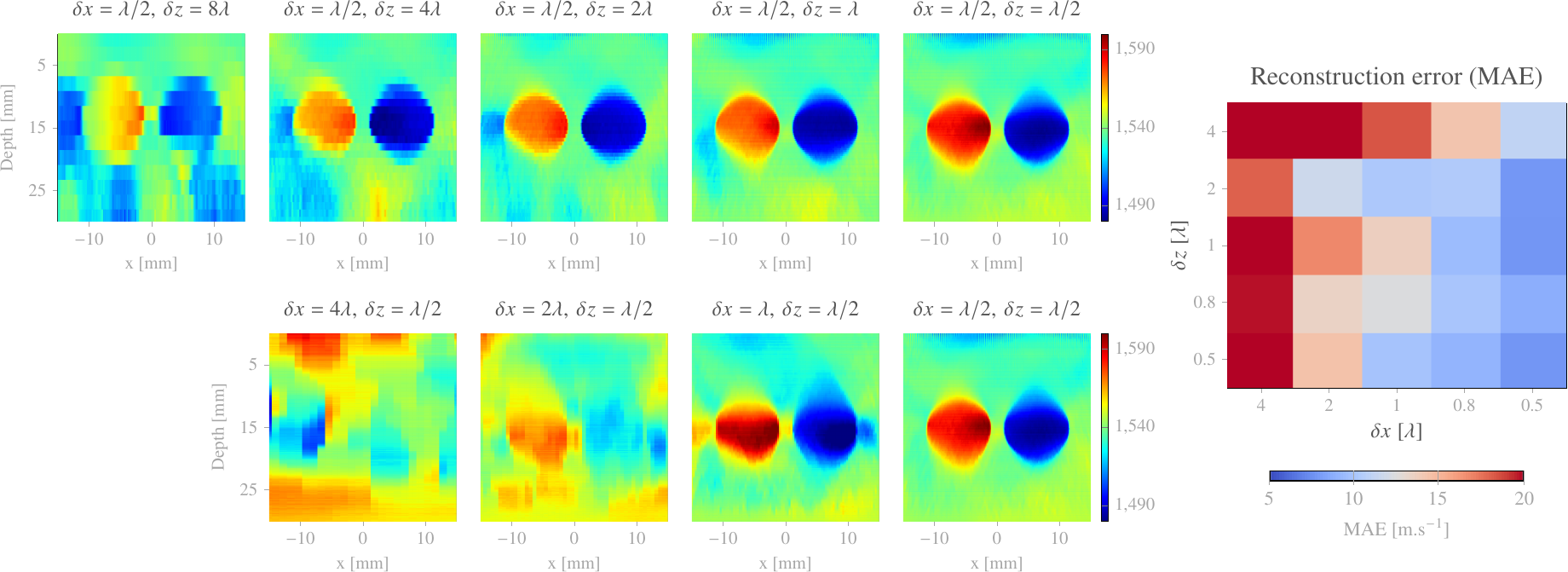}
	\caption{\textbf{Effect of reconstruction resolution}. Top left: Effect of axial resolution, for a lateral resolution of $\delta x = \lambda/2$. Bottom left: Effect of lateral resolution, for an axial resolution of $\delta z = \lambda/2$. Right: Mean absolute error for different lateral and axial resolutions.}\label{fig:sup_res}
\end{figure*}

	We observe that the lateral resolution has a noticeable effect on reconstruction accuracy and must be smaller than the wavelength in order to identify the two inclusions. In addition, shallow artifacts appear when the lateral resolution is smaller than the wavelength (and than the transducers pitch). Nevertheless, these artifacts are limited to the first wavelengths and the best reconstructed map is still the one obtained with $\delta x = \lambda / 2$.	
	Regarding the axial resolution, we observe a limited effect on the reconstruction accuracy. Axial resolutions of $2\lambda$ or even $4\lambda$ result in clearly identifiable inclusions, even though the $4\lambda$ case starts to show artifacts in depth.

This observation is in accordance with diffraction theory~\cite{wasik2026}, which predicts a scaling of the lateral resolution as $\delta x \sim \lambda / (2N\!A)$ and of the axial resolution as $\delta z \sim 2\lambda / N\!A^2$, with $\lambda$ the wavelength and $NA=\sin\alpha$ the numerical aperture that depends on the maximum observation angle $\alpha$. In this simulation, with an emission and reception directivity of approximately 35 degrees, we obtain $\delta x \sim \lambda/2$ and $\delta z \sim 6 \lambda$, which fits the limit cases of Supplementary Fig.~\ref{fig:sup_res}.

	This is a promising direction to reduce the computation load of wave velocity estimation. Nevertheless, the reported complexity of our method is in $\mathcal{O}\left(N_z N_x^2 \log N_x \right)$, or $\mathcal{O}\left(N_z N_x \log N_x\right)$ with the factorization \eqref{eq:fact2r}, so the axial resolution has a lower effect on complexity than lateral resolution.

\subsection{Optimization Parameters}\label{sup:optimization_parameters}

The last reconstruction parameters are the optimization parameters, including the learning rates, regularization weights and median filter size (last section in Table~\ref{tab:parameters}).

The median filter is required as demonstrated in Supplementary Section~\ref{sup:tv_filt}, but its size has no effect on the reconstructed wave velocity map. We keep it between 5 and 12 wavelengths, to avoid falling into local minima while allowing for recovering small details. In the 3D case, its size is kept to a minimum for computation time reasons.

	We study here the effect of the learning rate $\epsilon$ and regularization weight $\eta$. We limit our study to the learning of $\delta \sigma$, and suppose that it can be generalized to $\sigma_m$. As per the previous section, we reconstruct wave velocity maps for different pairs of $(\epsilon, \eta)$, and display the result in Supplementary Fig.~\ref{fig:sup_lr_reg}.
	Increasing the learning rate shows no big impact on the reconstructed velocity map, until it reaches a point where artifacts suddenly start to appear (see the case $\epsilon=10\cdot 10^{-3}$ and $\eta=5$), associated with a drop of the focusing quality at convergence (i.e. stuck a local optimum).
	On the other hand, tuning the regularization weight results in a slow transition between a map showing artifacts, especially in deeper regions, to a smooth map with an under-estimation of velocity variations in the inclusions at too high regularization weights.
	In addition to the effect on the wave velocity map, increasing $\epsilon$ or decreasing $\eta$ also reduces the number of iterations required to reach convergence and therefore the computation time.

\begin{figure*}[htp]
	\includegraphics[width=\textwidth]{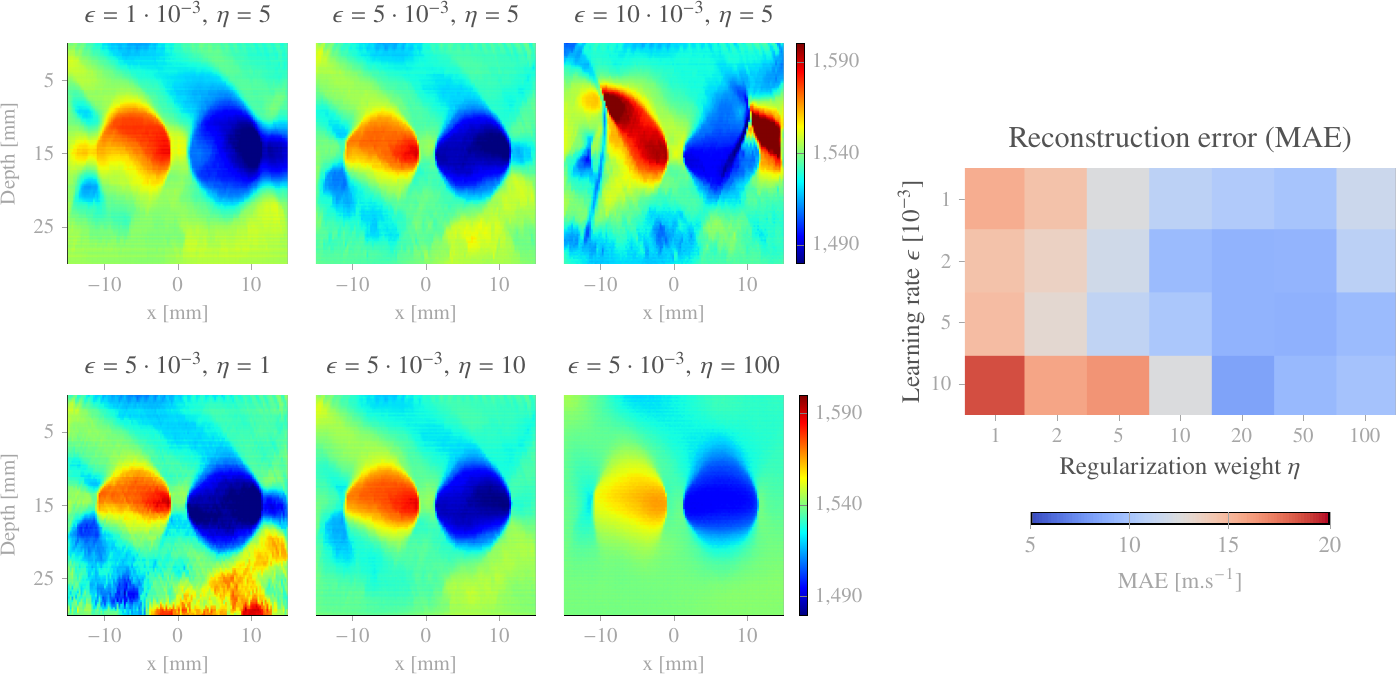}
	\caption{\textbf{Effect of learning rate $\epsilon$ and regularization weight $\eta$}. Top left: Effect of learning rate, for a regularization weight of $\eta=5$. Bottom left: Effect of regularization weight, for a learning rate of $\epsilon=5\cdot 10^3$. Right: Mean absolute error for different learning rates and regularization weights.}\label{fig:sup_lr_reg}
\end{figure*}

	The parameters proposed for the simulation case showed good results for all the cases studied in this work. However, we perform a case-by-case fine-tuning of these parameters to decrease computation time and optimize the results.
	This tuning is performed as follows. We first tune the regularization weight, choosing a trade-off between short-range artifacts and the risk of underestimation of velocity variations. We then increase slowly the learning rate until the velocity map at convergence changes, with the apparition of strongly spatially varying features (interpreted as artifacts) and a drop in the focusing quality.
	This tuning procedure can be repeated as the optimal regularization can slightly change with the learning rate, and reciprocally.
	We expect this tuning procedure to be necessary each time the acquisition parameters change (new frequencies, new probe, new emissions), and for each application case associated with changes in signal to noise ratio, velocity variation range and scale, and multiple scattering.

\section{Effect of Acquisition}\label{sec:acquisition_effect}

	In this section, we study the effect of the acquisition parameters, namely the wave emissions and the signal to noise ratio.

\subsection{Emissions}

	The acquisition cases studied in this work cover 1D and 2D canonical emissions, as well as plane wave emissions (breast tissue case).
	Thanks to the wavefield perspective of the Split-Step Fourier propagator, the proposed method is compatible with most emission strategies, especially the ones that can be described without using a reference wave velocity (e.g. with positions for canonical emissions or delay slopes for plane wave emissions).

	In the case of plane wave emissions, the number of emissions can be reduced in order to increase the achievable frame rate and reduce the data rate \cite{montaldo_coherent_2009}.
	In this section, we study the effect of the number of emissions and chosen angular range on the reconstructed wave velocity map to assess the compatibility of our method with fewer acquisitions.

\begin{figure*}[htp]
	\includegraphics[width=\textwidth]{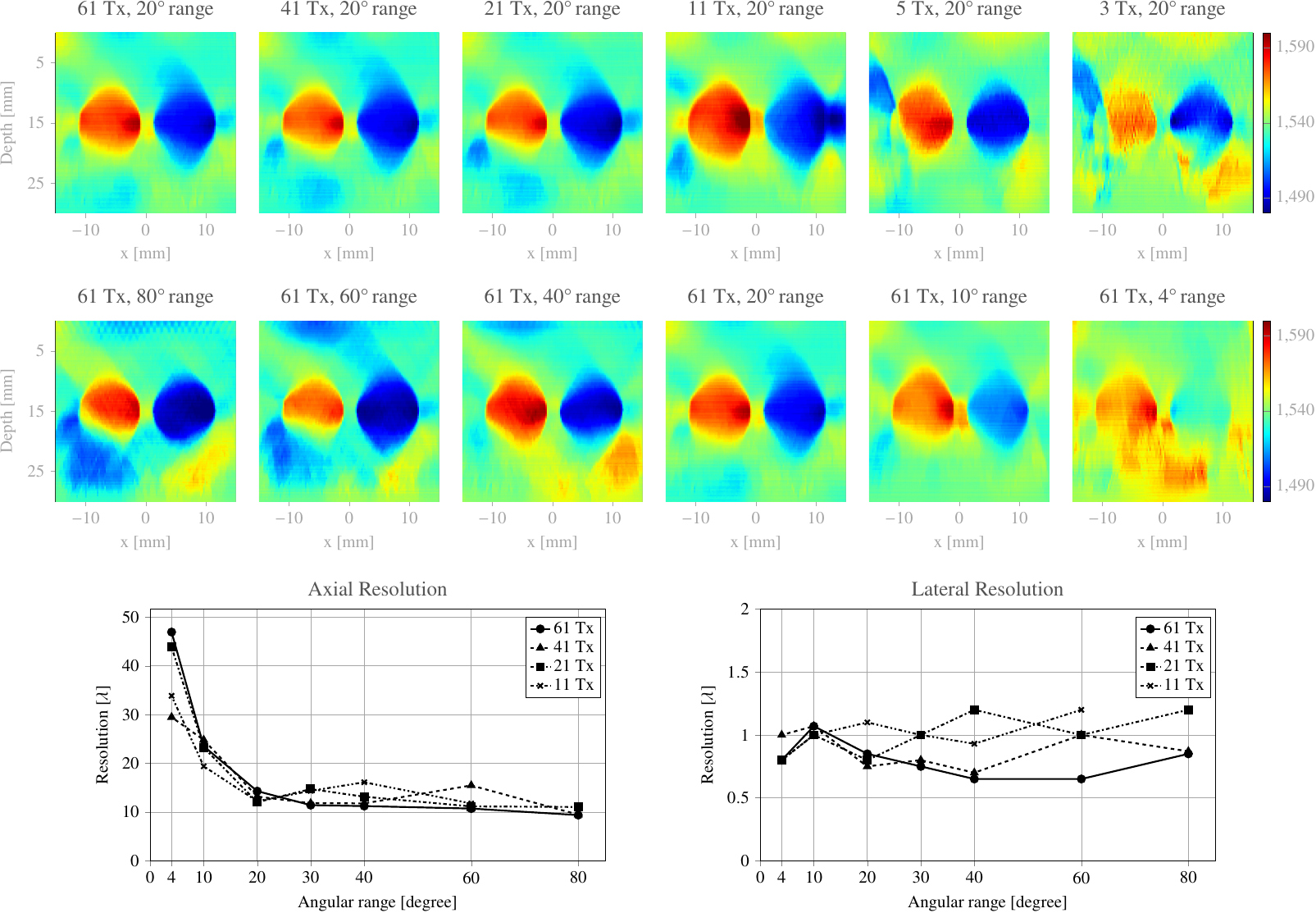}
	\caption{\textbf{Effect of number and angular range of emissions}. Top: Effect of the number of emissions, for an angular range of 20\textdegree\,(from -10\textdegree\, to 10\textdegree). Effect of the angular range for 61 emissions. Bottom left: Axial resolution of the reconstructed velocity map. Bottom right: Lateral resolution of the reconstructed velocity map.}\label{fig:sup_tx}
\end{figure*}

	Results displayed in Supplementary Fig.~\ref{fig:sup_tx} demonstrate that the proposed reconstruction can be performed with as few as 3 plane waves, and with an angular range as small as 10\textdegree . Decreasing the number of emissions increases the amount of artifacts in the reconstructed map but does not affect the shape nor the value of the velocity inclusions. On the other hand, reducing the angular range results in a reduced axial resolution of the reconstructed map and an underestimation of the wave variations.
	By measuring the axial and lateral velocity slope at the inclusion boundaries, we can estimate the axial and lateral resolution of the velocity map, computed as the estimated length required to reach a 50~m.s$^{-1}$ variation.
	We obtain a lateral resolution of $\sim1\lambda$, unaffected by the number of emissions and angular range, and an axial resolution of $\sim10\lambda$ which increases when reducing the angular range.

	We conclude that this method is theoretically compatible with sparse acquisitions. However, a reduced number of emissions is susceptible to lead to other limitations, such as a signal to noise ratio decrease, which can hinder wave velocity estimation in real-world configurations.

\subsection{Signal to Noise Ratio}

	To study the effect of presence of noise on the reconstructed velocity map, we consider a noiseless canonical simulation and add a gaussian complex noise $n$ to the raw data $s$, at different signal-to-noise ratio:
	\begin{equation}
		s_n(\mathbf{i}_\textrm{in}, \mathbf{o}_\textrm{out}, t) = s(\mathbf{i}_\textrm{in}, \mathbf{o}_\textrm{out}, t) + \alpha \cdot n(\mathbf{i}_\textrm{in}, \mathbf{o}_\textrm{out}, t),\quad\text{with}\; n(\mathbf{i}_\textrm{in}, \mathbf{o}_\textrm{out}, t)\sim\mathcal{N}\left(0, 1\right),
	\end{equation}
	where $s_n$ is the noisy signal and $i_{in}, o_{out}, t$ refer to the emission, reception and time.
	$\alpha$ is a prefactor computed as $\alpha=10^{-\text{snr}/20} \cdot \sigma_s$, with $\sigma_s$ the standard deviation of the noiseless signal and $\text{snr}$ the target signal to noise ratio (SNR).
	Reconstructed wave velocity maps are displayed in Supplementary Fig.~\ref{fig:sup_snr}.

\begin{figure*}[htp]
	\includegraphics[width=\textwidth]{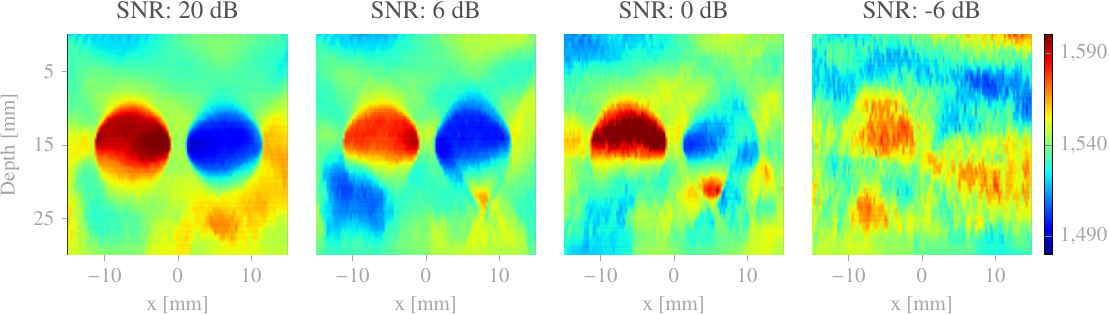}
	\caption{\textbf{Effect of signal to noise ratio}.}\label{fig:sup_snr}
\end{figure*}

	The presence of noise results in more artifacts, and the velocity inclusions are unidentifiable for negative signal to noise ratios.
	The presented method is therefore robust to noise up to $\sim$0~dB in the case of 128 single transducer emissions.

\section{Comparison with State of the Art Methods}\label{sec:sota}

\subsection{State of the Art Benchmark}

As mentioned in the introduction, several local speed of sound estimation techniques have been proposed in ultrasound in the past decade, based on straight-ray approximations or multi-layered geometries.
We separate these methods into two main classes.
A first type of method estimate the aberrations caused by wave velocity anomalies and recover a speed of sound map by inverting the tomographic problem to recover a speed of sound map \cite{stahli2020improved, stahli2023first}.
However, aberrations are estimated as phase perturbations to a constant velocity prior, limiting the application of such methods to small wave velocity contrasts.

	The second type of methods uses an iterative optimization process, in which aberrations are re-estimated for each wave velocity guess, enabling the recovery of velocity map with higher contrast \cite{ali2023sound,simson2025ultrasound}. Results of some of these methods, applied to the in vitro dataset available in \cite{ali2023sound}, are displayed in Supplementary Fig.~\ref{fig:sup_sota}, next to results of our method obtained on the same dataset.
	While all these methods are able to correct aberrations of the reflectivity image to some extent and to detect the positive velocity anomaly in the upper part of the medium, our method is the only one to render the circular inclusions.

\begin{figure*}[htp]
	\includegraphics[width=\textwidth]{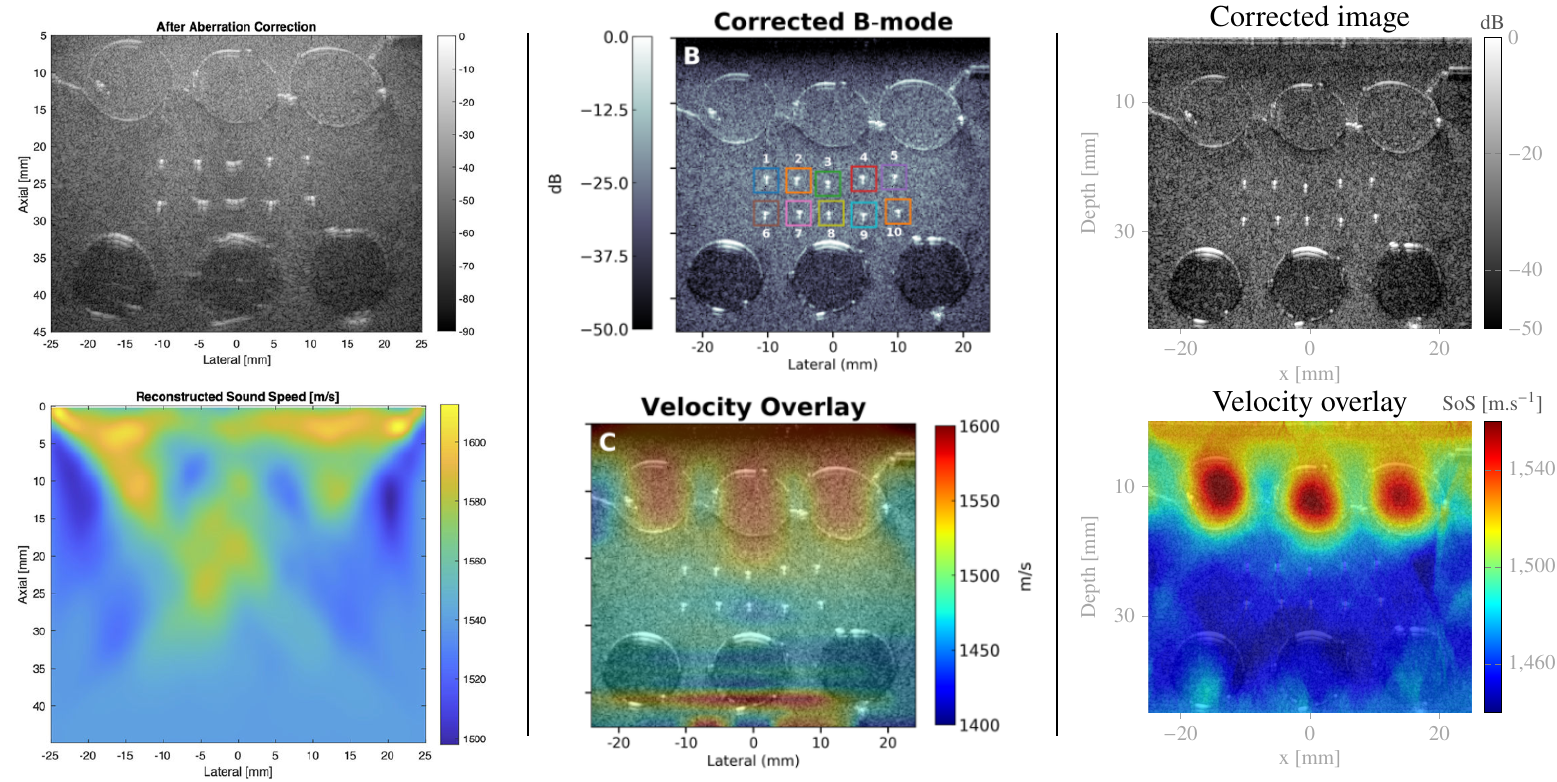}
	\caption{\textbf{Comparison between our method and state of the art methods on the in vitro experiment (from \cite{ali2023sound})}. Left: results presented in \cite{ali2023sound}. Middle: results presented in \cite{simson2025ultrasound}. Right: results from the proposed method.}\label{fig:sup_sota}
\end{figure*}

	To go further, the main contributions of our work lies in (i) a matrix imaging derived loss, able to probe the focusing quality in speckle even in strongly heterogeneous media; (ii) a wave-field propagation model, the Split-Step Fourier model, that efficiently models diffraction and refraction effects, implemented via the matrix imaging framework; (iii) an optimization scheme, in which local minima are studied through the matrix imaging perspective (using reflection matrices and RPSFs) and avoided via several optimization techniques.
	The combination of these three points allows the optimization of highly resolved velocity maps (with grid steps smaller than the wavelength), leading to velocity maps of high contrast and resolution.

	To further study the effect of each of these building blocks, we replace the loss and the propagation model with state-of-the-art alternatives and analyze the obtained velocity maps.

\subsection{Role of the Loss}

	We study 5 different state-of-the-art losses, namely the proposed focusing quality \eqref{eq:focusing_quality2}, the brightness, defined as the confocal energy (equal to \eqref{eq:focusing_quality2} without its normalization), the coherence factor defined in \eqref{eq:focusing_quality_d} and introduced in \cite{mallart1994adaptive}, the lag-one coherence introduced in \cite{long2018lag} and the common mid-point phase error (CMPPE) introduced in \cite{simson2025ultrasound}.

	To test these losses, we implement them as differentiable operators, and use them in place of the proposed focusing quality.
	Such an implementation is direct for the first four losses. The last one requires a deep modification of the proposed method as the CMPPE needs to track the origin of each wave reaching each point in the medium. This point of view is compatible with straight ray propagation operators, but less with wave-field propagation ones. Here, we implement the CMPPE by repeating the numerical focusing step for many sub-apertures in emission and reception, leading to computation times two order of magnitude higher than for other losses.
	Results are displayed in Supplementary Fig.~\ref{fig:sup_loss}.

\begin{figure*}[htp]
	\includegraphics[width=\textwidth]{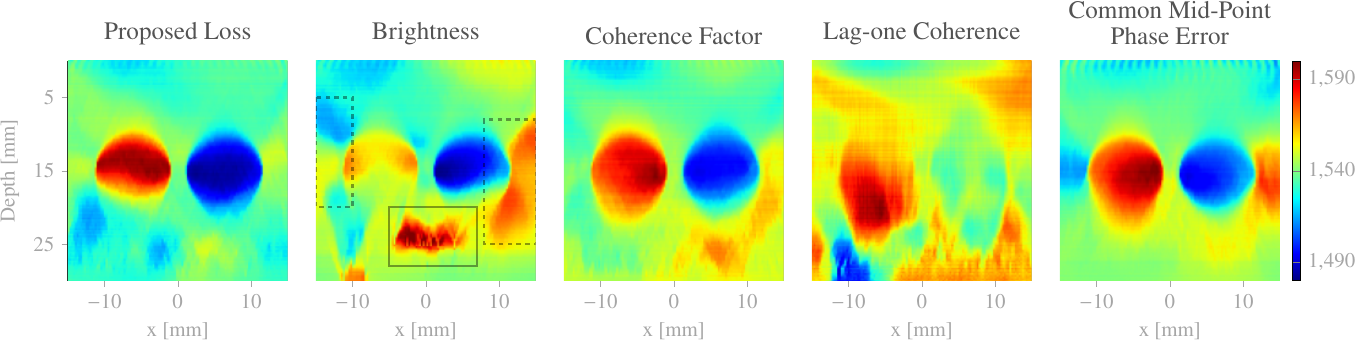}
	\caption{\textbf{Proposed method with various established losses}. In the brightness case, specific artifacts are emphasized with solid and dashed rectangles.}\label{fig:sup_loss}
\end{figure*}

	In the considered simulation, the proposed focusing quality, the coherence factor and the CMPPE show the best results, with small border artifacts in the CMPPE case.
	The brightness case shows strong artifacts at the borders and near the bright scatterers (solid rectangle). Indeed, removing the normalization in \eqref{eq:focusing_quality2} leads to a sensitivity to the medium reflectivity.
	Finally, the lag-one coherence results in a map of lower quality, even though the two inclusions can be identified qualitatively.

	This experiment shows that the proposed focusing quality and coherence factor are not the only quality criteria possible with our algorithm. Nevertheless, they are the ones that give the best results while being compatible with the split-step Fourier operator.

\subsection{Role of the Propagation Model}

	We challenge the need for an accurate wave-field propagation operator by switching it with a straight ray model. The latter supposes infinitely thin straight rays and therefore does not model refraction or diffraction effects. Results obtained after tuning the reconstruction parameters are displayed in Supplementary Fig.~\ref{fig:sup_model}.

\begin{figure}[htp]
	\includegraphics[width=0.5\textwidth]{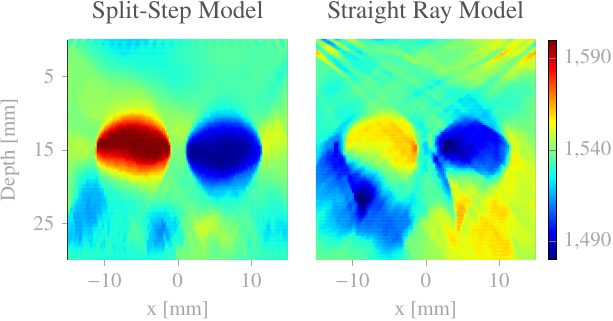}
	\caption{\textbf{Proposed method with two different propagation models}. Left: Split-Step Fourier propagation model (proposed). Right: Straight ray propagation model.}\label{fig:sup_model}
\end{figure}

	We observe that the straight ray model also renders the two velocity inclusions. However, it shows artifacts below these inclusions, that are likely caused by the inability to model the refraction and diffraction effects caused by the two inclusions. In addition, we see ray-like artifacts in the first 5~mm of the reconstructed map, and an underestimation of the velocity inside the inclusions.
	We conclude that a straight-ray model can qualitatively recover the wave velocity map but may result in artifacts and biases.

	It is worth mentioning that this experiment does not prove that a straight ray model is insufficient to recover accurate wave velocity maps. We solely show that the combination of our focusing quality with the split-step Fourier method and our optimization scheme enables the reconstruction of high-quality velocity map, and that switching the propagation model alone deteriorates the result.
	Another combination, more suited to the straight ray propagation, may yield better results.
	The theoretical validity of the straight ray model for velocity estimation is still an open question and needs to be studied mathematically. Advantageously, the use of the split-step Fourier method removes the need to answer this question.

\section{Computation Time and Memory Requirements}\label{sec:complexity}

	The computation complexity of the proposed method is largely driven by the split-step Fourier method, which scales in $\mathcal{O}\left(N_z N_x^2 \log N_x \right)$ or $\mathcal{O}\left(N_z N_{i} N_x \log N_x \right)$ in its factorized form, in 2D, with $N_z$ the number of points in the axial direction, $N_x$ in the lateral direction, and $N_i$ the number of emissions (cf Supplementary Section \ref{sup:fact}). In 3D, $N_x$ is replaced by $N_x N_y$.
	The computation of its derivative can be performed with a similar complexity, but requires to store a large amount of intermediate information.
	To reduce the memory footprint of the derivative computation, we propose to batch this processing along the frequencies (see Supplementary Section~\ref{sup:diff}).
	The size of the frequency batch acts therefore as a trade-off between computation time and memory requirements.
	For each imaging case presented in Table~\ref{tab:parameters}, we report in Supplementary Fig.~\ref{fig:sup_compute} the memory requirements and the computation time per iteration, obtained with a NVIDIA RTX 6000 Ada Generation GPU, for different frequency batch sizes.

\begin{figure}[htp]
	\includegraphics[width=0.8\textwidth]{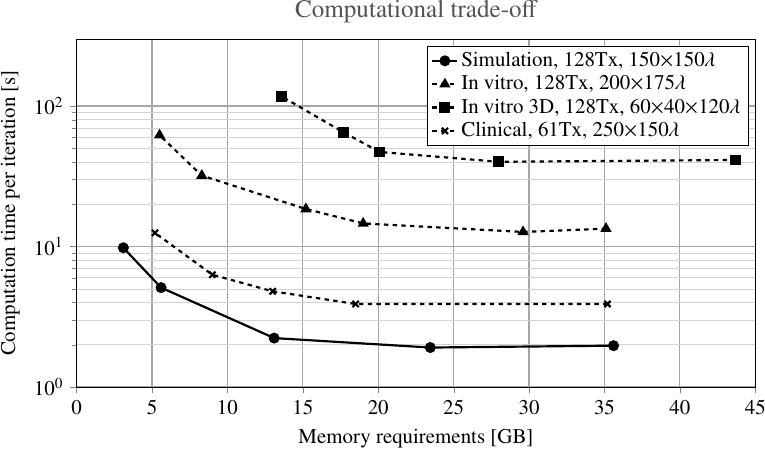}
	\caption{\textbf{Computation time and memory requirements for all experiments, with different batching size}. For each configuration, identified by the number of emissions and the grid size, different trade-offs between computation time and memory footprint can be reached by playing on the frequency batching size.}\label{fig:sup_compute}
\end{figure}

	This test is performed with the code provided together with this article, relying on PyTorch built-in GPU optimizations.
	A more precise GPU profiling shows that the memory requirements could be drastically reduced (of almost one order of magnitude) with a custom gradient computation implemented directly in CUDA.
	Similarly, we suspect that the computation time per iteration could be reduced using a pre-compilation and/or a custom GPU implementation.
	Nevertheless, we show here that all cases, even the 3D one, can be run on a GPU with a 16GB RAM memory, which is now common for the most recent GPU series.

\section{Extended Results}\label{sec:3d_extended}

\begin{figure*}[htp]
	\includegraphics[width=\textwidth]{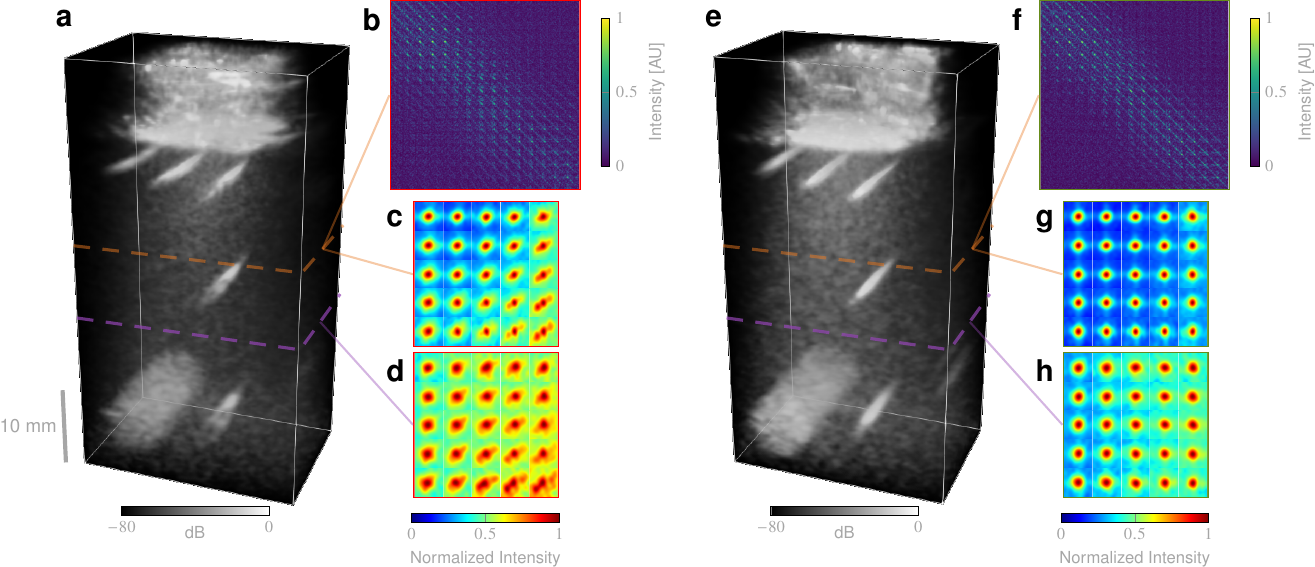}
	\caption{\textbf{Aberration correction of \textit{ex vivo} tissues in a 3D configuration}. \textbf{a}, aberrated reflection image obtained with a uniform speed of sound of 1540~m.s$^{-1}$. \textbf{b}, 4D Focused reflection matrix at 30~mm, flattened in 2D. \textbf{c-d} 2D RPSFs obtained from the antidiagonal profile of the 4D focused reflection matrices at 30 and 40~mm, respectivel. \textbf{e}, corrected reflection image. \textbf{f-h}, same as \textbf{b-d} but for the corrected case, showing a reduction of the size of the focal spots.}\label{fig:sup_fig_3_extended}
\end{figure*}

\begin{figure*}[htp]
	\includegraphics[width=\textwidth]{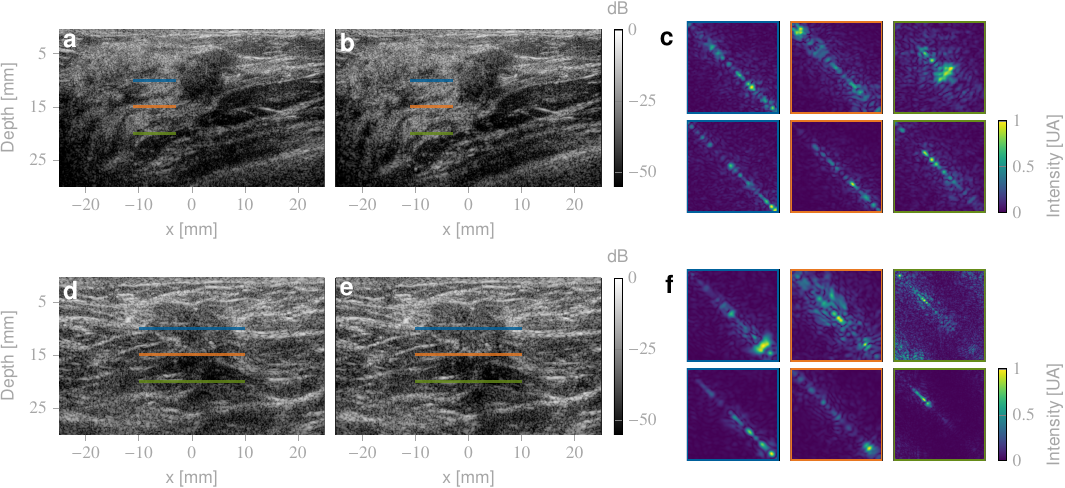}
	\caption{\textbf{Reflection matrix improvement for the breast tissue case}. \textbf{a, d}, aberrated reflection images obtained with a uniform speed of sound of 1480~m.s$^{-1}$. \textbf{b, e}, corrected reflection matrices. \textbf{c, f} focused reflection matrices, estimated at 10, 15 and 20~mm along the color lines. For both cases, top row shows the aberrated case, and bottom row shows the corrected case, highlighting the diagonal shape of the corrected matrices associated with focusing quality increases of 27 to 53\%.}\label{fig:sup_fig_4_extended}
\end{figure*}

\clearpage 

\paragraph*{Caption for Supplementary Video 1.}
 \textbf{Aberration compensation in the 2D \textit{in vitro} experiment.} The initial (red square) and corrected (green square) images (Fig.~\ref{fig:2d_invitro}b and e, respectively) are alternatively displayed to highlight the benefit of the optimized speed-of-sound map (Fig.~\ref{fig:2d_invitro}f) for the beamforming process and the resulting reflection image. The dynamic range of the B\&W scale is of 50 dB.\\
 
 \paragraph*{Caption for Supplementary Video 2.}
 \textbf{Aberration compensation of \textit{ex vivo} tissues in a 3D configuration.} The initial (red square) and corrected (green square, Fig.~\ref{fig:3d_invitro}b) images are alternatively displayed to highlight the benefit of the optimized speed-of-sound map (Fig.~\ref{fig:3d_invitro}d) for the beamforming process and the resulting reflection image. The dynamic range of the B\&W scale is of 80 dB.
 
  \paragraph*{Caption for Supplementary Video 3.}
 \textbf{Aberration compensation in the first breast tissue experiment.} The initial (red square) and corrected (green square, Fig.~\ref{fig:2d_invivo}b) images are alternatively displayed to highlight the benefit of the optimized speed-of-sound map (Fig.~\ref{fig:2d_invivo}c) for the beamforming process and the resulting reflection image. The dynamic range of the B\&W scale is of 55 dB.
 
   \paragraph*{Caption for Supplementary Video 4.}
 \textbf{Aberration compensation in the second breast tissue experiment.} The initial (red square) and corrected (green square, Fig.~\ref{fig:2d_invivo}d) images are alternatively displayed to highlight the benefit of the optimized speed-of-sound map (Fig.~\ref{fig:2d_invivo}e) for the beamforming process and the resulting reflection image. The dynamic range of the B\&W scale is of 55 dB.

   \paragraph*{Caption for Supplementary Video 5.}
 \textbf{Aberration compensation in the third breast tissue experiment.} The initial (red square) and corrected (green square, Fig.~\ref{fig:2d_invivo}f) images are alternatively displayed to highlight the benefit of the optimized speed-of-sound map (Fig.~\ref{fig:2d_invivo}g) for the beamforming process and the resulting reflection image. The dynamic range of the B\&W scale is of 55 dB.
 
    \paragraph*{Caption for Supplementary Video 6.}
 \textbf{Aberration compensation in the fourth breast tissue experiment.} The initial (red square) and corrected (green square, Fig.~\ref{fig:2d_invivo}h) images are alternatively displayed to highlight the benefit of the optimized speed-of-sound map (Fig.~\ref{fig:2d_invivo}i) for the beamforming process and the resulting reflection image. The dynamic range of the B\&W scale is of 55 dB.
 
     \paragraph*{Caption for Supplementary Video 7.}
 \textbf{Aberration compensation in the fifth breast tissue experiment.} The initial (red square) and corrected (green square, Fig.~\ref{fig:2d_invivo}j) images are alternatively displayed to highlight the benefit of the optimized speed-of-sound map (Fig.~\ref{fig:2d_invivo}k) for the beamforming process and the resulting reflection image. The dynamic range of the B\&W scale is of 55 dB.

\clearpage

\def\bibsection{\noindent \textbf{References.}}


\end{document}